\documentclass[showpacs,preprintnumbers,superscriptaddress,aps,nofootinbib,perprint,12pt]{revtex4}
\usepackage{amssymb}
\usepackage[centertags]{amsmath}
\usepackage{epsfig}
\usepackage{bm}
\usepackage{color}
\usepackage{graphicx,graphics}
\usepackage{multirow}
\usepackage{float}
\usepackage[pdfstartview=FitH]{hyperref}
\hypersetup{colorlinks=true, citecolor=blue,linkcolor=blue,filecolor=black,urlcolor=blue}
\allowdisplaybreaks[2]

\usepackage{slashed}

\begin{document}

\title{Higher twist contribution to fragmentation function in inclusive hadron production in $e^+e^-$ annihilation }

\author{Shu-yi Wei}
\affiliation{School of Physics, Shandong University, Jinan, Shandong 250100, China}

\author{Yu-kun Song}
\affiliation{Interdisciplinary Center for Theoretical Study and Department of Modern
Physics, University of Science and Technology of China, Anhui 230026, China}
\affiliation{Key Laboratory of Quark and Lepton Physics (CCNU), Ministry of Education, China}

\author{Zuo-tang Liang}
\affiliation{School of Physics, Shandong University, Jinan, Shandong 250100, China}

\begin{abstract}
We apply collinear expansion to inclusive hadron production in $e^+e^-$ annihilation and 
derive a formalism suitable for systematic study of leading as well as higher twist contributions to  fragmentation functions at the tree level. 
We make the calculations for hadrons with spin-$0$, spin-$1/2$ as well as spin-$1$ 
and obtain the results in terms of different components of fragmentation functions 
for the hadronic tensors, the differential cross section as well as hadron polarizations 
in different cases.  
The results show a number of interesting features such as the existence of 
transverse polarization for spin-$1/2$ hadrons at the twist-3 level, 
the quark polarization independence of the spin alignment of vector mesons. 
\end{abstract}

\maketitle

\newpage

\section{introduction}

Fragmentation function is one of the most important physical quantities in describing the hadron production in high energy reactions.
It quantifies the hadronization of quarks and/or gluons that occur in every high energy reaction process where hadron is produced and is therefore
a necessary ingredient in any complete description of processes involving hadron production.
The study of the fragmentation function provides not only such an important ingredient in describing high energy reactions
but also important information on the properties of Quantum Chromodynamics (QCD) and is therefore
a standing topic in the field of high energy physics.
Many progresses have been made and summarized in a number of recent reviews~\cite{Beringer:1900zz}.
Much attention has been attracted recently in particular 
in the spin dependence~\cite{Collins:1992kk,Jaffe:1993xb,Ji:1993vw,Chen:1994ar,Boer:1997mf,deFlorian:1997zj, Anselmino:2000vs, Bacchetta:2005rm,Boer:2008fr,Yuan:2009dw, Kanazawa:2013uia,Buskulic:1996vb, Ackerstaff:1997nh,Ackerstaff:1997kj, Ackerstaff:1997kd, Abreu:1997wd,ALEPH:2005ab,Abe:2005zx,Vossen:2011fk}.
This provides a new window to study fragmentation functions, to test hadronization models and to learn the properties of QCD.

Like parton distribution functions, parton fragmentation functions can be defined in terms of the quark and gluon field operators in a gauge invariant form.
The relationship between such gauge invariant fragmentation functions and the differential cross section is essential to the study 
of such fragmentation functions and to the description of high energy reactions. 
Such a relationship can be established using collinear expansion technique applied to the corresponding reaction.
To study the unpolarized reactions, collinear approximation is often valid to high accuracy and the leading twist contributions
are usually enough for the description of hadron production without polarizations.
This is in fact also the case in most of the current studies where only leading twist contributions are considered.
However,  it is unclear whether higher twist effects are also negligible in the polarized cases.
In particular, in the cases where transverse momentum is considered and the azimuthal asymmetry is studied,
such higher twist effects can be very important. It is therefore necessary and important to make 
a study including the leading and higher twist contributions in a systematic way.

The plan of this paper series is to make such a systematic study of higher twist effects in quark fragmentation processes.
In this paper, we start with inclusive hadron production in $e^+e^-$ annihilation at high energies.
We apply the collinear expansion technique to this process and present the formalism for calculating
leading and higher twist contributions in a consistent and systematic way.
We carry out the calculations up to twist-3 for spin-$1/2$ as well as spin-1 hadrons using this formalism. 
We present the results obtained for the hadronic tensors, the differential cross sections and 
the polarizations of hadrons in different cases. 
We also show how to proceed the calculations for contributions at twist-4 level and present 
the results for spin-$1/2$ particle production as an example. 
 
The rest of this paper is organized as follows. In Sec. II, we present the formalism for calculating leading and
higher twist contributions using collinear expansion technique.
In Sec. III, we carry out the calculations for the hadronic tensors for spin-0, spin-$1/2$ and spin-1 hadrons and present the 
corresponding results up to twist-3. 
In Sec. IV, we present the results for the differential cross sections and the polarizations of the hadrons. 
In Sec. V, we discuss the twist-4 contributions and present the results for spin-1/2 hadrons. 
We make a summary and give an outlook in Sec. VI.

\section{The formalism}

\begin{figure}[h!]
\includegraphics[width=0.5\textwidth]{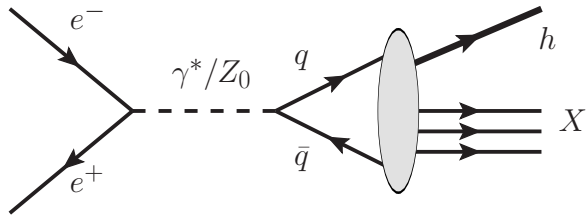}
\caption{Illustrating diagram for inclusive hadron production in $e^+e^-$ annihilation.}\label{ff1}
\end{figure}

We consider the inclusive hadron production process, $e^+e^-\to h+X$,  
as illustrated in Fig.~\ref{ff1}.
We use $l_1$ and $l_2$ to denote the 4-momenta of the incoming electron and positron,
and $q=l_1+l_2$ to denote the 4-momentum of the intermediate gauge boson. 
The momentum of the quark is denoted by $k$ and that of the produced hadron is denoted by $p$. 

For explicity, we consider $e^+e^-$ annihilation into hadrons either via electromagnetic interaction with the exchange of a virtual photon or via weak interaction with the exchange of a $Z^0$-boson. 
We do not consider the interference term and the results apply to reactions near the $Z^0$-pole where only the 
weak interaction term is considered or the energy is much lower than $Z^0$-mass where only electromagnetic interaction is needed. 
In this case, we get the differential cross section as given by,
\begin{equation}
d \sigma =  \frac{g_Z^4}{32s} L_{\mu'\nu'} (l_1, l_2) D_F^{\mu'\mu}(q) D_F^{\nu'\nu*}(q) W_{\mu\nu}(q,p,S) \frac{d^3p}{(2\pi)^2 2E_p}. \label{CS}
\end{equation}
Here $L^{\mu'\nu'}(l_1,l_2)$ is the leptonic tensor and for reactions with unpolarized leptons,
\begin{align}
  L_{\mu'\nu'}(l_1,l_2)=&\frac{1}{4} \mathrm{Tr}\left[\Gamma_{\mu'}^e \slashed l_1 \Gamma_{\nu'}^e \slashed l_2 \right].
\end{align}
where we use $\Gamma_{\mu'}^e$ instead of $\gamma_{\mu'}$ since the intermediate boson can be a photon or a $Z^0$-boson. 
In the case that the intermediate boson is a $Z^0$-boson (weak interaction), 
we have $\Gamma_{\mu'}^e = \gamma_{\mu'}(c_V^e - c_A^e \gamma^5)$ 
while $\Gamma_{\mu'}^e = \gamma_{\mu'}$ or equivalently $c_V=1$ and $c_A=0$ if it is a photon (electromagnetic interaction).
Correspondingly, the propagator is,
$D_{\mu'\mu} = (g_{\mu'\mu}-q_{\mu'}q_\mu /M_Z^2)/[(Q^2-M_Z^2)+i\Gamma_Z M_Z]$, 
and  $D_{\mu'\mu} = g_{\mu'\mu}/Q^2 $ respectively.
The weak coupling $g_Z=g/\cos\theta_W=e/\sin\theta_W\cos\theta_W$ where $e$ is the electron charge 
and $\theta_W$ is the Weinberg angle.
We note that, due to current conservation $q^\mu L_{\mu\nu}=0$, 
 the second part of $Z^0$ propagator does not contribute in this case. 
The leptonic tensor for $Z^0$-exchange is given by,
\begin{align}
L_{\mu'\nu'}(l_1,l_2)=&c_1^e\left[l_{1\mu'} l_{2\nu'}+l_{1\nu'} l_{2\mu'}-(l_1\cdot l_2)g_{\mu'\nu'}\right]
+ic_3^e\varepsilon_{\mu'\nu'\rho\sigma}l_{1}^{\rho}l_{2}^{\sigma},
\end{align}
where $c_1^e = (c_V^e)^2+(c_A^e)^2$ and $c_3^e = 2 c_V^e c_A^e$.

The hadronic tensor $W^{\mu\nu}$ is defined as,  
\begin{equation}
W^{\mu\nu}(q,p,S) = \frac{1}{2\pi}  \sum_X (2\pi)^4 \delta^4 (q-p - P_X) \langle 0| J^\nu (0) |p,S;X\rangle \langle p,S;X |J^\mu (0)|0\rangle. \label{ht001}
\end{equation}

\begin{figure}[h!]
\includegraphics[width=\textwidth]{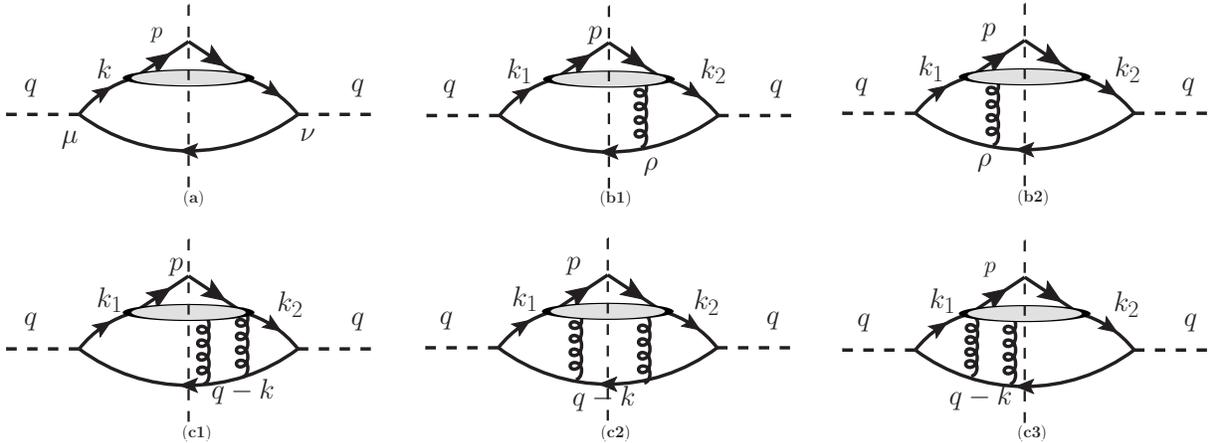}
\caption{The first few Feynman diagrams as examples of the diagram series with exchange of $j$ gluon(s).
In (a), (b) and (c), we see the case for $j=0$, $1$ and $2$ respectively. 
The gluon momentum in (b) is $k_1-k_2$, while in (c), they are $k-k_1$ and $k_2-k$ respectively.} \label{fdmg}
\end{figure}

To the leading order, the hadronic tensor is shown in Fig.~\ref{fdmg}(a), and is given by,
\begin{align}
W_{\mu\nu}^{(0)} (q,p,S) = \int \frac{d^4k}{(2\pi)^4} \mathrm{Tr}\left[ \hat{H}_{\mu\nu}^{(0)} (k,q) \hat{\Pi}^{(0)}(k,p,S) \right]. \label{eq:W0}
\end{align}
It is given by a trace of the calculable hard part, 
\begin{equation}
\hat{H}_{\mu\nu}^{(0)} (k,q) = \Gamma_\mu^q (\slashed q - \slashed k) \Gamma_\nu^q  (2\pi) \delta_+ \left((q-k)^2\right), \label{hp001}
\end{equation}
and the matrix element defined by,
\begin{equation}
\hat{\Pi}^{(0)} (k,p,S)= \frac{1}{2\pi} \sum_X\int d^4\xi e^{-ik\xi} \langle 0 |\psi(0)|hX\rangle \langle hX| \bar\psi(\xi)|0\rangle.
\end{equation}
Here, as well as in the following of this paper, unless explicitly stated, a summation over the 
quark flavor and color is implicit and the flavor index is omitted.  
In fact, the hard part as that given in Eq.~(\ref{hp001}) is independent of quark color so that the summation over color 
leads simply to a color factor $N_c=3$. This should be included in the final result of the cross section. 
Also, we use quark as an example for explicity. 
All the expressions can be extended to include anti-quark contributions. 
We use the same forms for the expressions so that we can simply include the anti-quark contributions 
by extending the sum over flavors to anti-quarks as well. 
There is no essential difference between the calculations for quarks and those for anti-quarks. 
The results are similar and we will specify if there is any difference in the corresponding places 
in the following of this paper.

It is well known that, because the two quark fields in the matrix element $\hat{\Pi}^{(0)}$ do not share the same space-time coordinate, 
$\hat{\Pi}^{(0)}$ is not local (color) gauge invariant. 
To get the gauge invariant form, we need to consider the final-state interaction in QCD, 
and apply the collinear expansion technique~\cite{Ellis:1982wd,Qiu:1990xxa}. 
The collinear expansion was first applied to deeply inelastic lepton-nucleon scattering (DIS) and 
provides an unique way to obtain a consistent formalism that relates the gauge invariant parton 
distribution and/or correlation functions to the measurable quantities such as the 
differential cross section 
including leading as well as higher twist contributions.  
It has been recently extended to semi-inclusive DIS with nucleon and nucleus targets 
for jet production\cite{Liang:2006wp,Liang:2008vz,Gao:2010mj,Song:2010pf,Song:2013sja} 
and corresponding expressions for the azimuthal asymmetries and nuclear dependences have been obtained.  
It is therefore also necessary to apply collinear expansion to $e^+e^-$ annihilation to obtain the 
corresponding formalism in order to establish the relationship between the differential cross section and the fragmentation functions. 
We now summarize the main steps and results in the following. 

\subsection{Gauge invariance and collinear expansion} 

To get the gauge invariant form for the fragmentation function in $e^+e^-$ annihilation, 
we need to consider the multiple gluon scattering similar to those considered in 
deep inelastic scattering \cite{Ellis:1982wd}. 
In this case, we need to consider the diagrams with exchange of $j=1,2,\ldots$ gluon(s) between 
the blob and the lower Fermion line in Fig.\ref{fdmg}(a).    
As examples, we show those with exchange of one and two gluons in Figs.~\ref{fdmg}(b) and (c).

Taking such multiple gluon scattering into account, the hadronic tensor is given by, 
\begin{equation}
W_{\mu\nu}=\sum_{j,c}W_{\mu\nu}^{(j,c)}=W_{\mu\nu}^{(0)}+W_{\mu\nu}^{(1,L)}+W_{\mu\nu}^{(1,R)}+\ldots, 
\end{equation}
where we use the superscript to denote the contribution from the Feynman diagram with 
exchange of $j=0,1,2,\ldots$ gluon(s) and $c$ denotes the position of the cut line which takes  
$L$ or $R$ for $j=1$, $c=L$, $M$ or $R$ for $j=2$ and corresponds to 
Fig.~\ref{fdmg}(b1), (b2), (c1), (c2) and (c3) respectively.
For the case with one gluon exchange, we have, 
\begin{align}
W_{\mu\nu}^{(1,c)}(q,p,S)=& \int \frac{d^4k_1}{(2\pi)^4} \frac{d^4k_2}{(2\pi)^4} \mathrm{Tr}[ \hat H^{(1,c)\rho}_{\mu\nu} (k_1,k_2,q)  \hat \Pi_{\rho}^{(1,c)}(k_1,k_2,p,S) ], \label{eq:W1} 
\end{align}
where $c=L$ or $R$ and the hard parts are given by,
\begin{align}
\hat H^{(1,L)\rho}_{\mu\nu}(k_1,k_2,q) = & \Gamma_\mu^q (\slashed q - \slashed k_1) \gamma^\rho \frac{\slashed k_2- \slashed q}{(k_2-q)^2-i\epsilon}  \Gamma^q_\nu  (2\pi) \delta_+\left((q-k_1)^2\right),\\
\hat H^{(1,R)\rho}_{\mu\nu}(k_1,k_2,q) = & \Gamma_\mu^q \frac{\slashed k_1 - \slashed q}{(k_1-q)^2+i\epsilon} \gamma^\rho (\slashed q - \slashed k_2) \Gamma^q_\nu  (2\pi) \delta_+ \left((q-k_2)^2\right), 
\end{align}
and the soft matrices are defined as,
\begin{align}
\hat \Pi_{\rho}^{(1,L)}(k_1,k_2,p,S) = & \frac{1}{2\pi} \sum_X\int d^4 \xi d^4 \eta e^{-ik_1\xi} e^{-i(k_2-k_1)\eta} \langle 0 | g A_\rho(\eta) \psi(0) |hX \rangle \langle hX | \bar \psi(\xi)|0\rangle,\label{eq:Pi1L}\\
\hat \Pi_{\rho}^{(1,R)}(k_1,k_2,p,S) = & \frac{1}{2\pi} \sum_X\int d^4 \xi d^4 \eta e^{-ik_1\xi} e^{-i(k_2-k_1)\eta} \langle 0 |\psi(0) |hX \rangle \langle hX | \bar \psi(\xi) g A_\rho(\eta) |0\rangle.\label{eq:Pi1R}
\end{align}
For $j=2$, the corresponding results are shown by Figs.~\ref{fdmg}(c1), (c2) and (c3), and are given by, 
\begin{align}
W_{\mu\nu}^{(2,c)}(q,p,S)=& \int \frac{d^4k_1}{(2\pi)^4} \frac{d^4k_2}{(2\pi)^4} \frac{d^4k}{(2\pi)^4} 
\mathrm{Tr}[ \hat H^{(2,c)\rho\sigma}_{\mu\nu} (k_1,k,k_2,q)  \hat \Pi_{\rho\sigma}^{(2,c)}(k_1,k,k_2,p,S) ],
\end{align}
where $c=L$, $M$ or $R$, and the hard parts are given by,
\begin{align}
 &\hat H_{\mu\nu}^{(2,L)\rho\sigma} (k_1,k,k_2, q) =  \Gamma_\mu^q (\slashed q - \slashed k_1) \gamma^\rho \frac{\slashed k- \slashed q}{(k-q)^2-i\epsilon} \gamma^\sigma  \frac{\slashed k_2- \slashed q}{(k_2-q)^2-i\epsilon}  \Gamma^q_\nu  (2\pi) \delta_+\left((q-k_1)^2\right),\\
 &\hat H_{\mu\nu}^{(2,M)\rho\sigma} (k_1,k,k_2, q) =  \Gamma_\mu^q \frac{\slashed k_1- \slashed q}{(k_1-q)^2+i\epsilon} \gamma^\rho (\slashed q - \slashed k) \gamma^\sigma  \frac{\slashed k_2- \slashed q}{(k_2-q)^2-i\epsilon}  \Gamma^q_\nu  (2\pi) \delta_+\left((q-k)^2\right),\\
  &\hat H_{\mu\nu}^{(2,R)\rho\sigma} (k_1,k,k_2, q) = \Gamma_\mu^q \frac{\slashed k_1- \slashed q}{(k_1-q)^2+i\epsilon} \gamma^\rho \frac{\slashed k- \slashed q}{(k-q)^2+i\epsilon} \gamma^\sigma  (\slashed q - \slashed k_2) \Gamma^q_\nu  (2\pi) \delta_+\left((q-k_2)^2\right),
\end{align}
and the soft matrices are difined as,
\begin{align}
\hat \Pi_{\rho\sigma}^{(2,L)}(k_1,k,k_2,p,S) = & \frac{1}{2\pi} \sum_X\int d^4 \xi d^4 \eta_1 d^4 \eta_2 e^{-ik_1\xi} e^{-i(k-k_1)\eta_1} e^{-i(k_2-k)\eta_2}  \nonumber\\
& \times \langle 0 | g A_\rho(\eta_1) gA_\sigma (\eta_2) \psi(0) |hX \rangle \langle hX | \bar \psi(\xi)|0\rangle,\\
\hat \Pi_{\rho\sigma}^{(2,M)}(k_1,k,k_2,p,S) = & \frac{1}{2\pi} \sum_X\int d^4 \xi d^4 \eta_1 d^4 \eta_2 e^{-ik_1\xi} e^{-i(k-k_1)\eta_1} e^{-i(k_2-k)\eta_2} \nonumber\\
& \times \langle 0 | gA_\sigma (\eta_2) \psi(0) |hX \rangle \langle hX | \bar \psi(\xi)g A_\rho(\eta_1) |0\rangle,\\
\hat \Pi_{\rho\sigma}^{(2,R)}(k_1,k,k_2,p,S) = & \frac{1}{2\pi} \sum_X\int d^4 \xi d^4 \eta_1 d^4 \eta_2 e^{-ik_1\xi} e^{-i(k-k_1)\eta_1} e^{-i(k_2-k)\eta_2} \nonumber\\
& \times \langle 0 |\psi(0) |hX \rangle \langle hX | \bar \psi(\xi) g A_\rho(\eta_1) gA_\sigma (\eta_2) |0\rangle,
\end{align}

We note that none of such soft matrices is local (color) gauge invariant. 
To get the gauge invariant form, we need to apply the collinear expansion 
as proposed in \cite{Ellis:1982wd}, which is carried out in the following four steps as summarized in~\cite{Liang:2006wp}. 

(1) Make a Taylor expansion of all the hard parts around $k_i = p/z_i $, e.g., 
\begin{align}
\hat H_{\mu\nu}^{(0)}(k,q) = & \hat H_{\mu\nu}^{(0)} (z) + \frac{\partial \hat H_{\mu\nu}^{(0)}(z)}{\partial k_\rho} \omega_\rho^{\ \rho'} k_{\rho'} + \frac{1}{2} \frac{\partial^2 \hat H_{\mu\nu}^{(0)}(z)}{\partial k_\rho \partial k_\sigma} \omega_\rho^{\ \rho'} k_{\rho'} \omega_{\sigma}^{\ \sigma'} k_{\sigma'}+\cdots , \\
\hat H_{\mu\nu}^{(1,L)\rho}(k_1,k_2,q) = & \hat H_{\mu\nu}^{(1,L)\rho} (z_1,z_2) + 
\frac{\partial \hat H_{\mu\nu}^{(1,L)\rho}(z_1,z_2)}{\partial k_{1\sigma}} \omega_\sigma^{\ \sigma'} k_{1\sigma'} + 
\frac{\partial \hat H_{\mu\nu}^{(1,L)\rho}(z_1,z_2)}{\partial k_{2\sigma}} \omega_\sigma^{\ \sigma'} k_{2\sigma'} +\cdots ,
\end{align}
where, different from that for deeply inelastic scattering~\cite{Liang:2006wp}, for the fragmentation process, $z_i$ is defined as $z_i=p^+/k^+_i$.  
The momentum of the hadron is taken as $p=p^+\bar n$ i.e. we use the light cone coordinate and 
take the direction of motion of the hadron as $z$-direction.  
The unit vectors in this coordinate system are denoted by $\bar n$, $n$ and $n_\perp$. 
In $e^+e^-$-annihilation, we choose the lepton plane as $xoz$-plane and the transverse component of 
the momentum of the incident electron is taken as the $x$-direction, and that of incident positron is in the minus $x$-direction.  
The projection operator $\omega_\rho^{\ \rho'}$ is defined as 
$\omega_\rho^{\ \rho'} \equiv g_\rho^{\ \rho'} -\bar n_\rho n^{\rho'}$.
We also use the short notations such as $\hat H_{\mu\nu}^{(0)}(z)\equiv \hat H_{\mu\nu}^{(0)}(k,q)|_{k=p/z}$, 
$\partial\hat H_{\mu\nu}^{(0)}(z)/\partial k_\rho\equiv \partial \hat H_{\mu\nu}^{(0)}(k,q)/\partial k_\rho|_{k=p/z}$ and so on.

Here, as usual in the collinear expansion, we neglect the $n$-component of the hadron momentum. This component should take the form $(M^2/2p^+)n$, where $M$ is the hadron mass.  Compared to the $\bar n$-component, it is suppressed by a factor $(M/p^+)^2$ and contributes only at twist-4 level. This was discussed in the past in e.g.~\cite{Nachtmann:1973mr, Georgi:1976ve} and we will also come back to this point in Sec. V where examples of twist-4 contributions are given.   

(2) Decompose the gluon fields into longitudinal and transverse components, i.e.,
\begin{equation}
A_\rho(y) = A^+(y) \bar n_\rho + \omega_\rho^{\ \rho'} A_{\rho'}(y).
\end{equation}

(3) Apply the Ward identities such as,
\begin{align}
& \frac{\partial \hat H^{(0)}_{\mu\nu}(z)}{\partial k_\rho}=  - \hat H^{(1,L)\rho}_{\mu\nu} (z,z) - \hat H^{(1,R)\rho}_{\mu\nu}(z,z),\\
& \frac{\partial \hat H^{(1,L)\rho}_{\mu\nu}(z_1,z_2)}{\partial k_{1,\sigma}}=  - \hat H^{(2,L)\rho\sigma}_{\mu\nu} (z_1,z_1,z_2) - \hat H^{(2,M)\rho\sigma}_{\mu\nu}(z_1,z_1,z_2),\\
& \frac{\partial \hat H^{(1,L)\rho}_{\mu\nu}(z_1,z_2)}{\partial k_{2,\sigma}}=  - \hat H^{(2,R)\rho\sigma}_{\mu\nu} (z_1,z_2,z_2) ,\\
&p_\rho \hat H^{(1,L)\rho}_{\mu\nu}(z_1,z_2) = - \frac{z_1 z_2}{z_2-z_1-i\epsilon} H^{(0)}_{\mu\nu} (z_1),\\
&p_\rho \hat H^{(1,R)\rho}_{\mu\nu}(z_1,z_2) = - \frac{z_1 z_2}{z_1-z_2+i\epsilon} H^{(0)}_{\mu\nu} (z_2),\\
&p_\rho \hat H^{(2,L)\rho\sigma}_{\mu\nu}(z_1,z,z_2) = - \frac{z_1 z}{z-z_1-i\epsilon} H^{(1,L)\sigma}_{\mu\nu} (z_1,z_2). 
\end{align}

(4) Add all the terms with the same hard part together and we obtain the hadronic tensor in the gauge invariant form as given by, 
\begin{equation}
W_{\mu\nu} = \sum_{j,c}\tilde{W}^{(j,c)}_{\mu\nu}=\tilde{W}^{(0)}_{\mu\nu}+\tilde{W}^{(1,L)}_{\mu\nu}+\tilde{W}^{(1,R)}_{\mu\nu}+\cdots, 
\end{equation}
where the tilded $W$'s are given by,
\begin{align}
&\tilde{W}_{\mu\nu}^{(0)}(q,p,S) = \int \frac{dk^+}{2\pi p^+} \mathrm{Tr}\left[ \hat H^{(0)}_{\mu\nu}(z) \hat \Xi^{(0)} (z,p,S;n) \right],\label{eq:Wtilde0}\\
&\tilde{W}_{\mu\nu}^{(1,L)}(q,p,S)= \int \frac{dk_1^+}{2\pi p^+}\frac{dk_2^+}{2\pi p^+}  \mathrm{Tr} \left[ \hat H^{(1,L)\rho}_{\mu\nu}(z_1,z_2) \omega_\rho^{\ \rho'} \hat \Xi_{\rho'}^{(1,L)} (z_1,z_2,p,S;n) \right],\label{eq:Wtilde1L}\\
&\tilde{W}_{\mu\nu}^{(1,R)}(q,p,S)= \int \frac{dk_1^+}{2\pi p^+}\frac{dk_2^+}{2\pi p^+} \mathrm{Tr} \left[ \hat H^{(1,R)\rho}_{\mu\nu}(z_1,z_2) \omega_\rho^{\ \rho'} \hat \Xi_{\rho'}^{(1,R)} (z_1,z_2,p,S;n) \right],\label{eq:Wtilde1R} \\
&\tilde{W}_{\mu\nu}^{(2,L)}(q,p,S)= \int \frac{dk_1^+}{2\pi p^+}\frac{dk_2^+}{2\pi p^+}\frac{dk^+}{2\pi p^+}
  \mathrm{Tr} \left[ \hat H^{(2,L)\rho\sigma}_{\mu\nu}(z_1,z,z_2) \omega_\rho^{\ \rho'}\omega_\sigma^{\ \sigma'} \hat \Xi_{\rho'\sigma'}^{(2,L)} (z_1,z,z_2,p,S;n) \right],\label{eq:Wtilde2L}\\
&\tilde{W}_{\mu\nu}^{(2,M)}(q,p,S)= \int \frac{dk_1^+}{2\pi p^+}\frac{dk_2^+}{2\pi p^+}\frac{dk^+}{2\pi p^+}
  \mathrm{Tr} \left[ \hat H^{(2,M)\rho\sigma}_{\mu\nu}(z_1,z,z_2) \omega_\rho^{\ \rho'}\omega_\sigma^{\ \sigma'} \hat \Xi_{\rho'\sigma'}^{(2,M)} (z_1,z,z_2,p,S;n) \right],\label{eq:Wtilde2M}\\
&\tilde{W}_{\mu\nu}^{(2,R)}(q,p,S)= \int \frac{dk_1^+}{2\pi p^+}\frac{dk_2^+}{2\pi p^+}\frac{dk^+}{2\pi p^+}
  \mathrm{Tr} \left[ \hat H^{(2,R)\rho\sigma}_{\mu\nu}(z_1,z,z_2) \omega_\rho^{\ \rho'}\omega_\sigma^{\ \sigma'} \hat \Xi_{\rho'\sigma'}^{(2,R)} (z_1,z,z_2,p,S;n) \right]. \label{eq:Wtilde2R}
\end{align}
Here, the new correlator $\hat\Xi^{(j)}$'s are given by,
\begin{align}
&\hat \Xi^{(0)} (z,p,S;n)=\sum_X \int \frac{p^+d \xi^-}{2\pi} e^{-ik^+\xi^-} 
   \langle 0 | \mathcal{L}^\dagger(0,\infty)\psi(0) |hX\rangle \langle hX| \bar\psi(\xi^-) \mathcal{L}(\xi^-,\infty) |0\rangle, \label{nnlsp001}\\
&\hat \Xi^{(1,L)}_{\rho} (z_1,z_2,p,S;n) = \sum_X\int  \frac{p^+d\xi^- p^+d\eta^-}{2\pi} e^{-ik_1^+\xi^--i(k_2^+-k_1^+)\eta^-}  \nonumber\\
&\phantom{XXXXXXX}\times \langle 0 | \mathcal{L}^\dagger (\eta^-,\infty) D_{\rho}(\eta^-) \mathcal{L}^\dagger (0,\eta^-) \psi(0) |hX\rangle \langle hX| \bar\psi(\xi^-) \mathcal{L}(\xi^-,\infty) |0\rangle, \label{eq:Xi1L}\\
&\hat \Xi^{(1,R)}_{\rho} (z_1,z_2,p,S;n)  = \sum_X\int \frac{ p^+d\xi^- p^+d\eta^-}{2\pi} e^{-ik_1^+\xi^--i(k_2^+-k_1^+)\eta^-}  \nonumber\\
& \phantom{XXXXXXX} \times \langle 0 | \mathcal{L}^\dagger (0,\infty)\psi(0) |hX\rangle \langle hX| \bar\psi(\xi^-) \mathcal{L}(\xi^-,\eta^-) {\overleftarrow{D}}_{\rho}(\eta^-) \mathcal{L}(\eta^-,\infty) |0\rangle, \label{eq:Xi1R}\\
&\hat \Xi^{(2,L)}_{\rho\sigma}  (z_1,z,z_2,p,S;n) =  \sum_X\int  \frac{p^+d\xi^- p^+d\eta_1^- p^+d\eta_2^-}{2\pi} e^{-ik_1^+\xi^--i(k^+-k_1^+)\eta_1^- -i(k_2^+-k^+)\eta_2^-}  \nonumber\\
& \phantom{XX}\times \langle 0 | \mathcal{L}^\dagger (\eta_1^-,\infty) D_{\rho}(\eta_1^-) \mathcal{L}^\dagger (\eta_2^-,\eta_1^-) D_{\sigma}(\eta_2^-) \mathcal{L}^\dagger (0,\eta_2^-) \psi(0) |hX\rangle 
\langle hX| \bar\psi(\xi^-) \mathcal{L}(\xi^-,\infty) |0\rangle, \label{twist4l} \\
&\hat \Xi^{(2,M)}_{\rho\sigma}  (z_1,z,z_2,p,S;n) =  \sum_X\int  \frac{p^+d\xi^- p^+d\eta_1^- p^+d\eta_2^-}{2\pi} e^{-ik_1^+\xi^--i(k^+-k_1^+)\eta_1^- -i(k_2^+-k^+)\eta_2^-}  \nonumber\\
& \phantom{XX} \times\langle 0 | \mathcal{L}^\dagger (\eta_2^-,\infty) D_{\sigma}(\eta_2^-) \mathcal{L}^\dagger (0,\eta_2^-) \psi(0) |hX\rangle 
\langle hX| \bar\psi(\xi^-) \mathcal{L}(\xi^-,\eta_1^-) D_{\rho}(\eta_1^-)\mathcal{L}(\eta_1^-,\infty)  |0\rangle, \\
&\hat \Xi^{(2,R)}_{\rho\sigma} (z_1,z,z_2,p,S;n) = \sum_X\int  \frac{p^+d\xi^- p^+d\eta_1^- p^+d\eta_2^-}{2\pi} e^{-ik_1^+\xi^--i(k^+-k_1^+)\eta_1^- -i(k_2^+-k^+)\eta_2^-}  \nonumber\\
&  \phantom{XX} \times\langle 0 | \mathcal{L}^\dagger (0,\infty) \psi(0) |hX\rangle 
 \langle hX| \bar\psi(\xi^-) \mathcal{L}(\xi^-,\eta_1^-) D_{\rho}(\eta_1^-)\mathcal{L}(\eta_1^-,\eta_2^-) D_{\sigma}(\eta_2^-)  \mathcal{L} (\eta_2^-,\infty) |0\rangle, \label{twist4r}
\end{align}
where $D_{\rho} (\eta)= -i\partial_\rho + gA_{\rho}(\eta)$ is the covariant derivative,  
and the gauge link $\mathcal{L}$ is given by the following path integral,
\begin{align}
\mathcal{L}(\xi^-,\infty) &=Pe^{ig\int_{\xi^-}^\infty d\eta^-A^+(\eta_-)} \nonumber\\
&= 1 + ig\int_{\xi^-}^\infty d \eta^- A^+(\eta^-) + (ig)^2 \int_{\xi^-}^\infty d \eta_1^- \int_{\xi^-}^{\eta_1^-} d \eta_2^- A^+(\eta_2^-)A^+(\eta_1^-) + \cdots,
\end{align}
which guarantees the correlation matrices gauge invariant.

The hard parts in the $\tilde W^{(j)}$'s such as those given by Eqs.~(\ref{eq:Wtilde0})-(\ref{eq:Wtilde2R}) depend 
only on the longitudinal momentum fractions of the quarks and they are given by, 
\begin{align}
&\hat H^{(0)}_{\mu\nu} (z) = \Gamma_\mu^q \left(\slashed q - \slashed p /z \right) \Gamma^q_\nu (2\pi) \delta_+\left((q-p/z)^2\right), \label{hard00}\\
&\hat H^{(1,L)\rho}_{\mu\nu} (z_1,z_2) = \Gamma_\mu^q \left( \slashed q  - \slashed p/z_1 \right) \gamma^\rho
\frac{\slashed p/z_2 - \slashed q}{(p/z_2 - q)^2-i\epsilon} \Gamma^q_\nu (2\pi) \delta_+\left( (q-p/z_1)^2\right), 
\label{hard01L}\\
&\hat H^{(1,R)\rho}_{\mu\nu} (z_1,z_2) = \Gamma^q_\mu \frac{\slashed p/z_1 - \slashed q}{(p/z_1 - q)^2+i\epsilon} \gamma^\rho \left( \slashed q  - \slashed p /z_2\right)  \Gamma^q_\nu   (2\pi) \delta_+\left( (q-p/z_2)^2 \right), \label{hard01R} \\
&\hat H^{(2,L)\rho\sigma}_{\mu\nu} (z_1,z,z_2) = \Gamma_\mu^q \left( \slashed q  - \slashed p /z_1 \right) \gamma^\rho
\frac{\slashed p/z - \slashed q}{(p/z - q)^2-i\epsilon} \gamma^\sigma \frac{\slashed p/z_2 - \slashed q}{(p/z_2 - q)^2-i\epsilon}  \Gamma^q_\nu  \nonumber\\
& \phantom{XXXXXXXXXXXXXXXX}  \times (2\pi) \delta_+\left( (q-p/z_1)^2\right), \label{hard02L}\\
&\hat H^{(2,M)\rho\sigma}_{\mu\nu} (z_1,z,z_2) = \Gamma_\mu^q \frac{\slashed p/z_1 - \slashed q}{(p/z_1 - q)^2+i\epsilon} \gamma^\rho \left( \slashed q  - \slashed p/z \right) \gamma^\sigma \frac{\slashed p/z_2 - \slashed q}{(p/z_2 - q)^2-i\epsilon}  \Gamma^q_\nu  \nonumber\\
&  \phantom{XXXXXXXXXXXXXXXX} \times(2\pi) \delta_+\left( (q-p/z)^2\right), \label{hard02M}\\
&\hat H^{(2,R)\rho\sigma}_{\mu\nu} (z_1,z,z_2) = \Gamma_\mu^q \frac{\slashed p/z_1 - \slashed q}{(p/z_1 - q)^2+i\epsilon} \gamma^\rho 
\frac{\slashed p/z - \slashed q}{(p/z - q)^2+i\epsilon} \gamma^\sigma \left( \slashed q  - \slashed p/z_2 \right)  \Gamma^q_\nu  \nonumber\\
&  \phantom{XXXXXXXXXXXXXXXX} \times (2\pi) \delta_+\left( (q-p/z_2)^2\right). \label{hard02R}
\end{align}
We will refer to them as the collinear-expanded hard parts in the following of this paper.
We also note that $\hat H^{(1,L)\rho}_{\mu\nu} (k_1,k_2,q)=\gamma^0H^{(1,R)\rho\dag}_{\nu\mu} (k_2,k_1,q)\gamma^0$,   
$\hat H^{(2,L)\rho\sigma}_{\mu\nu} (k_1,k,k_2,q)=\gamma^0H^{(2,R)\sigma\rho\dag}_{\nu\mu} (k_2,k,k_1,q)\gamma^0$, 
and 
$\tilde{W}_{\mu\nu}^{(1,L)}(q,p,S)=\tilde{W}_{\nu\mu}^{(1,R)*}(q,p,S)$, 
$\tilde{W}_{\mu\nu}^{(2,L)}(q,p,S)=\tilde{W}_{\nu\mu}^{(2,R)*}(q,p,S)$. 
We give the expressions for both of them for symmetry.

We emphasize that all the results given by Eqs.(\ref{eq:Wtilde0})-(\ref{eq:Wtilde2R}) are derived from the series of diagrams such as those shown in Fig.2 
following the four steps for the collinear expansion described above. 
As in \cite{Ellis:1982wd},  we have carried out the derivations up to the second order in $g^2$ and have been convinced that such derivations
can be extended to even higher orders if needed. 
The gauge links inside the correlators $\hat\Xi^{(j)}$'s are obtained in the derivations without any arbitrariness. 
For example, all the first terms in the expansion of the $W^{(j)}$'s with $A^+$ component of gluon field are summed together 
to give the $\tilde W^{(0)}$ where all the corresponding terms containing $A^+$ and $j>0$ go to the gauge link. 
The first derivative term in the expansion with $\omega_\rho^{\ \rho'} k_{\rho'}$ is converted to $\omega_\rho^{\ \rho'} \partial_{\rho'}$ 
and combines  with the $\omega_\rho^{\ \rho'} A_{\rho'}(y)$ term to form the covariant derivative and so on.
We see also clearly why the projection operator $\omega_\rho^{\ \rho'}$ exists in  $\tilde W^{(j)}$ for $j>0$. 

We also emphasize that using collinear expansion we obtain the hadronic tensor as a sum of the $\tilde W$'s. 
Each of these $\tilde W^{(j)}$ 's receives contributions from all the infinite number of diagrams in the diagram series 
as illustrated by Fig. 2.   
The contributions from this diagram series are re-organized by using collinear expansion so that 
the correlators have the gauge invariant forms given by Eqs.(\ref{nnlsp001}-\ref{twist4r}). 
We should note that the contributions of these $\tilde W^{(j)}$'s to the hadronic tensor 
contain the leading and higher twists as well and can be calclulated order by order. 
The leading contribution in each $\tilde W^{(j)}$ is twist-$(2+j)$.  
We should also note that,  when going to twist-4 or higher, there are also contributions from other diagrams 
that are not included in this diagram series. 
To make a complete study in that case, we need also to take those contributions into account.
In this paper, we concentrate only on the results from this series of diagrams but specify clearly 
if more diagrams should be taken into account for a complete calculation for the specified case. 

\subsection{Simplifying the Results\label{str}}

Another very nice feature of the results is that, because the collinear-expanded hard parts 
given by Eqs.~(\ref{hard00})-(\ref{hard02R}) contain only the longitudinal components of the quark momenta 
and also due to the presence of the projection operator $\omega_\rho^{\ \rho'}$ in the cases for $j>0$, 
these results can be simplified in a great deal. 
The collinear-expanded hard parts, multiplied by the projection operator(s) $\omega_\rho^{\ \rho'}$ 
for $j>0$, can e.g. be simplified into,   
\begin{align}
&\hat H_{\mu\nu}^{(0)} (z) = z_B^2\pi \hat h^{(0)}_{\mu\nu} \delta (z-z_B), \\
&\hat H_{\mu\nu}^{(1,L)\rho} (z_1,z_2) \omega_\rho^{\ \rho'} 
=-\frac{\pi z_B^2}{2p\cdot q} \hat h_{\mu\nu}^{(1)\rho} \delta (z_1-z_B) \omega_\rho^{\ \rho'}, \\
&\hat H_{\mu\nu}^{(1,R)\rho} (z_1,z_2) \omega_\rho^{\ \rho'} 
=-\frac{\pi z_B^2}{2p\cdot q} \gamma_0 \hat h_{\nu\mu}^{(1)\rho\dagger} \gamma_0 \delta (z_2-z_B) \omega_\rho^{\ \rho'}, \\
&\hat H_{\mu\nu}^{(2,M)\rho\sigma} (z_1,z,z_2) \omega_\rho^{\ \rho'} \omega_\sigma^{\ \sigma'} 
= \frac{2\pi z_B^2}{(2 p\cdot q)^2} \hat h_{\mu\nu}^{(2)\rho\sigma} \delta (z-z_B) \omega_\rho^{\ \rho'} \omega_\sigma^{\ \sigma'} ,\\
&\hat H_{\mu\nu}^{(2,L)\rho\sigma} (z_1,z,z_2) \omega_\rho^{\ \rho'} \omega_\sigma^{\ \sigma'} 
= \frac{2\pi z_B^2}{(2p \cdot q)^2} \Bigl(p^\sigma \hat h^{(1)\rho}_{\mu\nu} - \frac{z_2 z_B \hat N^{(2)\rho\sigma}_{\mu\nu}}{z_2-z_B - i\epsilon} \Bigr)  \delta (z_1-z_B) \omega_\rho^{\ \rho'} \omega_\sigma^{\ \sigma'} ,\\
&\hat H_{\mu\nu}^{(2,R)\rho\sigma} (z_1,z,z_2) \omega_\rho^{\ \rho'} \omega_\sigma^{\ \sigma'} 
=\frac{2\pi z_B^2}{(2p \cdot q)^2} \Bigl(p^\sigma \hat h^{(1)\rho\dagger}_{\nu\mu} - \frac{z_1 z_B \hat N^{(2)\rho\sigma\dagger}_{\nu\mu}}{z_1-z_B + i\epsilon} \Bigr) 
 \delta (z_2-z_B) \omega_\rho^{\ \rho'} \omega_\sigma^{\ \sigma'},
\end{align}
where $z_B\equiv 2p\cdot q /Q^2 $, 
$\hat h^{(0)}_{\mu\nu} = \Gamma_\mu^q \slashed n \Gamma_\nu^q/p^+ $, 
$\hat h_{\mu\nu}^{(1)\rho} = \Gamma_\mu^q \slashed n \gamma^\rho \slashed{\bar n}\Gamma_\nu^q$,
$\hat N^{(2)\rho\sigma}_{\mu\nu} = q^- \Gamma_\mu \gamma^\rho \slashed n \gamma^\sigma \Gamma_\nu$, 
and $\hat h_{\mu\nu}^{(2)\rho\sigma}=p^+ \Gamma_\mu \slashed{\bar n} \gamma^\rho \slashed n \gamma^\sigma \slashed{\bar n} \Gamma_\nu/2$.
We see in particular the following two features: 
(1) the $z$-dependence of the collinear-expanded hard parts, 
multiplied by the projection operator $\omega_\rho^{\ \rho'}$ for $j>0$, 
is usually very simple and in particular for $j=0$ and $1$,  it is contained only in the $\delta$-function; 
(2) these hard parts for $j>0$ usually depend on less number of parton momenta compared 
to the corresponding hard parts before the collinear expansion. 
For example, $\hat H_{\mu\nu}^{(1,c)\rho}(z_1,z_2)\omega_\rho^{\ \rho'}$ 
depends only on one parton momentum (eithor $z_1$ or $z_2$);  
$\hat H_{\mu\nu}^{(2,M)\rho}(z_1,z,z_2)\omega_\rho^{\ \rho'} \omega_\sigma^{\ \sigma'}$ 
depends only on $z$, while $\hat H_{\mu\nu}^{(2,L)\rho}(z_1,z,z_2)\omega_\rho^{\ \rho'} \omega_\sigma^{\ \sigma'} $ 
or $\hat H_{\mu\nu}^{(2,R)\rho}(z_1,z,z_2)\omega_\rho^{\ \rho'} \omega_\sigma^{\ \sigma'} $ 
each has two terms one of which depends only on either $z_1$ or $z_2$, the other depends on both $z_1$ and $z_2$ 
but none of them depends on $z$. 
This implies that we can carry out the integration over some of the parton momenta in the corresponding 
correlators and simplify the expressions for the hadronic tensors. The results are given by, 
\begin{align}
\tilde{W}_{\mu\nu}^{(0)}(q,p,S) =&  \frac{1}{2} \mathrm{Tr}\left[ \hat h^{(0)}_{\mu\nu} \hat \Xi^{(0)} (z_B,p,S;n) \right], \label{lwsht}\\
\tilde{W}_{\mu\nu}^{(1,L)}(q,p,S)=& -\frac{1}{4p\cdot q} 
\mathrm{Tr} \left[ \hat h^{(1)\rho}_{\mu\nu}\omega_\rho^{\ \rho'} \hat \Xi_{\rho'}^{(1)} (z_B,p,S;n) \right],\\
\tilde{W}_{\mu\nu}^{(1,R)}(q,p,S)=& -\frac{1}{4p\cdot q}
\mathrm{Tr} \left[ \hat h^{(1)\rho\dag}_{\nu\mu}\omega_\rho^{\ \rho'} \hat \Xi_{\rho'}^{(1)\dag} (z_B,p,S;n) \right],\\
\tilde{W}_{\mu\nu}^{(2,M)}(q,p,S)=& \frac{1}{4(p\cdot q)^2} \mathrm{Tr} \left[ \hat h_{\mu\nu}^{(2)\rho\sigma}  \omega_\rho^{\ \rho'} \omega_\sigma^{\ \sigma'} \hat \Xi_{\rho'\sigma'}^{(2A)} (z_B,p,S;n)  \right], \\
\tilde{W}_{\mu\nu}^{(2,L)}(q,p,S)=& \frac{1}{4(p\cdot q)^2} \mathrm{Tr}\left[ \hat h_{\mu\nu}^{(1)\rho} \omega_\rho^{\ \rho'} \hat \Xi_{\rho'}^{(2B)} (z_B,p,S;n) + \hat N_{\mu\nu}^{(2)\rho\sigma}  \omega_\rho^{\ \rho'} \omega_\sigma^{\ \sigma'} \hat \Xi_{\rho'\sigma'}^{(2C)} (z_B,p,S;n) \right],\\
\tilde{W}_{\mu\nu}^{(2,R)}(q,p,S)=& \frac{1}{4(p\cdot q)^2} \mathrm{Tr}\left[ \hat h_{\nu\mu}^{(1)\rho \dagger} \omega_\rho^{\ \rho'} \hat \Xi_{\rho'}^{(2B)\dagger} (z_B,p,S;n) 
+ \hat N_{\nu\mu}^{(2)\rho\sigma\dagger}  \omega_\rho^{\ \rho'} \omega_\sigma^{\ \sigma'} \hat \Xi_{\rho'\sigma'}^{(2C)\dagger} (z_B,p,S;n) \right], \label{t4sht}
\end{align}
where one correlator $\hat \Xi^{(1)}_{\rho}$ and 
three correlators $\hat \Xi^{(2A)}_{\rho\sigma}$, $\hat \Xi^{(2B)}_{\rho}$ and $\hat \Xi^{(2C)}_{\rho\sigma}$ 
 are involved for $\tilde W^{(1)}$ and $\tilde W^{(2)}$ repectively and they are defined as,
\begin{align}
&\hat \Xi^{(1)}_{\rho} (z_B,p,S;n)\equiv \frac{1}{2\pi} \int d(1/z_2) \hat \Xi^{(1,L)}_{\rho} (z_B,z_2,p,S;n), \\
&\hat \Xi^{(2A)}_{\rho\sigma} (z_B,p,S;n)\equiv \frac{1}{(2\pi)^2} \int d(1/z_1) d (1/z_2) \hat \Xi^{(2,M)}_{\rho\sigma} (z_1,z_B,z_2,p,S;n),\\
&\hat \Xi^{(2B)}_{\rho} (z_B,p,S;n)\equiv \frac{1}{(2\pi)^2} \int d(1/z) d(1/z_2)  p^\sigma \hat \Xi^{(2,L)}_{\rho\sigma} (z_B,z,z_2,p,S;n),\\
&\hat \Xi^{(2C)}_{\rho\sigma} (z_B,p,S;n)\equiv \frac{1}{(2\pi)^2} \int d(1/z) d(1/z_2) \frac{z_2z_B}{z_2-z_B-i\epsilon} \hat \Xi^{(2,L)}_{\rho\sigma} (z_B,z,z_2,p,S;n).
\end{align}
The corresponding field operator expressions are, 
\begin{align}
 \hat \Xi^{(1)}_{\rho} (z,p,S;n) = &
 \sum_X\int  \frac{p^+d\xi^-}{2\pi} e^{-ip^+\xi^-/z}  \nonumber\\  
&\times\langle 0 | \mathcal{L}^\dagger (0,\infty) D_{\rho}(0) \psi(0) |hX\rangle \langle hX| \bar\psi(\xi^-) \mathcal{L}(\xi^-,\infty) |0\rangle, \label{t3sp} \\ 
 \hat \Xi^{(2A)}_{\rho\sigma} (z,p,S;n)= & 
\sum_X \int  \frac{p^+d\xi^-}{2\pi} e^{-ip^+\xi^-/z}  \nonumber\\ 
& \times \langle 0 | \mathcal{L}^\dagger (0,\infty) D_{\rho}(0) \psi(0) |hX\rangle 
\langle hX| \bar\psi(\xi^-) D_\sigma (\xi^-)  \mathcal{L}(\xi^-,\infty) |0\rangle, \\
\hat \Xi^{(2B)}_{\rho} (z,p,S;n)= & 
\sum_X \int  \frac{p^+d\xi^-}{2\pi} e^{-ip^+\xi^-/z}  \nonumber\\ 
&\times p^\sigma\langle 0 | \mathcal{L}^\dagger (0,\infty) D_{\rho}(0)  D_\sigma (0) \psi(0) |hX\rangle 
\langle hX|\bar\psi(\xi^-) \mathcal{L}(\xi^-,\infty) |0\rangle,\\
 \hat \Xi^{(2C)}_{\rho\sigma} (z,p,S;n)= & 
\sum_X\int  \frac{p^+d\xi^- p^+d\eta^-}{2\pi} \frac{dz_2}{2\pi} \frac{1}{z_2^2} \frac{z_2 z}{z_2-z - i\epsilon} e^{-ip^+\xi^-/z-ip^+\eta^-/z_2}\nonumber \\
&\times \langle 0 | \mathcal{L}^\dagger (\eta^-,\infty) D_{\rho}(\eta^-) D_{\sigma} (\eta^-) 
 \mathcal{L}^\dagger (0,\eta^-)  \psi(0) |hX\rangle \langle hX| \bar\psi(\xi^-) \mathcal{L}(\xi^-,\infty) |0\rangle. \label{t4sp}
\end{align}

We note once more that, because the hard part $\hat H_{\mu\nu}^{(1,R)\rho} (z_1,z_2) \omega_\rho^{\ \rho'}$ depends only 
on one of the two parton momenta, we can carry out the integration over the other and obtain the correlator $\Xi^{(1)}_\rho$ that depends 
only on the corresponding one parton momemtum. 
In terms of the field operators, this implies that the gluon field or the covariant derivative is at the same space-time point as 
the quark (or anti-quark) field. 
Similar for the cases with even higher $j$. 
Again, such results are derived in this systematic formulation using collinear expansion.


To proceed further, we expand the involved matrices $\hat\Xi$'s in terms of $\gamma$-matrices. 
Since both $\hat h^{(0)}_{\mu\nu}$, $\hat h^{(1)\rho}_{\mu\nu}$, $\hat h_{\mu\nu}^{(2)\rho\sigma}$ 
and $\hat N^{(2)\rho\sigma}_{\mu\nu}$ all have odd number of $\gamma$-matrices, 
only $\gamma_\alpha$ and $\gamma_5\gamma_\alpha$ terms in the expansions of the $\hat\Xi$'s 
contribute. For example, for $j=0$ and $1$, we denote,
\begin{align}
\hat \Xi^{(0)} (z,p,S;n) = & \Xi^{(0)}_\alpha (z,p,S;n) \gamma^\alpha 
      + \tilde \Xi^{(0)}_\alpha (z,p,S;n) \gamma_5\gamma^\alpha+..., \\
\hat \Xi^{(1)}_\rho (z,p,S;n) = & \Xi^{(1)}_{\rho \alpha}(z,p,S;n) \gamma^\alpha 
      + \tilde \Xi^{(1)}_{\rho\alpha} (z,p,S;n) \gamma_5\gamma^\alpha+... , 
\end{align}
and obtain the hadronic tensors as, 
\begin{align}
&\tilde{W}_{\mu\nu}^{(0)}(q,p,S) =  \frac{1}{2} \left\{
  \mathrm{Tr}\left[\hat h^{(0)}_{\mu\nu} \gamma^\alpha\right]\Xi^{(0)}_\alpha (z_B,p,S;n) +
  \mathrm{Tr}\left[\hat h^{(0)}_{\mu\nu} \gamma_5\gamma^\alpha\right] \tilde\Xi^{(0)}_\alpha (z_B,p,S;n)\right\}, \label{ltht}\\
&\tilde{W}_{\mu\nu}^{(1,L)}
= -\frac{1}{4p\cdot q}\left\{ 
  \mathrm{Tr} \left[ \hat h^{(1)\rho}_{\mu\nu}\gamma^\alpha\right]
          \omega_\rho^{\ \rho'} \Xi_{\rho'\alpha}^{(1)} (z_B,p,S;n) +
 \mathrm{Tr} \left[ \hat h^{(1)\rho}_{\mu\nu}\gamma_5\gamma^\alpha\right]
          \omega_\rho^{\ \rho'} \tilde\Xi_{\rho'\alpha}^{(1)} (z_B,p,S;n)  \right\},\\
&\tilde{W}_{\mu\nu}^{(1,R)}
  = -\frac{1}{4p\cdot q}\left\{
  \mathrm{Tr} \left[ \hat h^{(1)\rho\dag}_{\nu\mu}\gamma^{\alpha\dag}\right]
          \omega_\rho^{\ \rho'}  \Xi_{\rho'\alpha}^{(1)*} (z_B,p,S;n) +
  \mathrm{Tr} \left[ \hat h^{(1)\rho\dag}_{\nu\mu}(\gamma_5\gamma^\alpha)^\dag\right]
          \omega_\rho^{\ \rho'}  \tilde\Xi_{\rho'\alpha}^{(1)*} (z_B,p,S;n) \right\}. \label{t3rchd}
\end{align}
The involved matrix elements are given by,
\begin{align}
&\Xi^{(0)}_\alpha (z,p,S;n) =  \sum_X\int \frac{p^+d \xi^-}{8\pi} e^{-ik^+\xi^-} \mathrm{Tr} \left[
\gamma_\alpha\langle 0 |  \mathcal{L}^\dagger(0,\infty) \psi(0) |hX\rangle 
\langle hX |\bar\psi(\xi^-) \mathcal{L}(\xi^-,\infty) |0\rangle \right], \label{xi0}\\
&\tilde\Xi^{(0)}_\alpha (z,p,S;n) =\sum_X\int \frac{p^+d \xi^-}{8\pi} e^{-ik^+\xi^-} \mathrm{Tr} \left[
\gamma_\alpha\gamma_5\langle 0 |  \mathcal{L}^\dagger(0,\infty) \psi(0) |hX\rangle 
\langle hX |\bar\psi(\xi^-) \mathcal{L}(\xi^-,\infty) |0\rangle \right], \label{xi05}\\
&\Xi^{(1)}_{\rho\alpha} (z,p,S;n)=\sum_X\int \frac{p^+d \xi^-}{8\pi} e^{-ik^+\xi^-} \mathrm{Tr} \left[
\gamma_\alpha \langle 0 | \mathcal{L}^\dagger(0,\infty) D_{\rho}(0)\psi(0) |hX\rangle 
\langle hX |\bar\psi(\xi^-) \mathcal{L}(\xi^-,\infty) |0\rangle \right], \label{xi1}\\
&\tilde\Xi^{(1)}_{\rho\alpha} (z,p,S;n) =  \sum_X\int \frac{p^+d \xi^-}{8\pi} e^{-ik^+\xi^-} \mathrm{Tr} \left[
\gamma_\alpha\gamma_5 \langle 0 | \mathcal{L}^\dagger(0,\infty) D_{\rho}(0)\psi(0) |hX\rangle 
\langle hX |\bar\psi(\xi^-) \mathcal{L}(\xi^-,\infty) |0\rangle \right].\label{xi15}
\end{align}
They are Lorentz vectors and tensors of second rank with different behaviors under space reflection respectively. 
We note in particular that, as can be seen from Eqs.~(\ref{xi0}-\ref{xi15}), 
 the dimension of $\Xi^{(0)}_\alpha$ or $\tilde\Xi^{(0)}_\alpha$ is 1 
 while that for $\Xi^{(1)}_{\rho\alpha}$ or $\tilde\Xi^{(1)}_{\rho\alpha}$ is 2. 
 This is important when we analyse the Lorentz structure of them in terms of 
 the 4-vectors $p$, $n$ and so on. 
 Also because of parity invariance, we have
\begin{align}
&\Xi^{(0)}_{\alpha} (z,p,S;n) = \Xi^{(0) \alpha} (z, \tilde p, -\tilde S; \tilde n),\label{eq:Parity1}\\
&\tilde\Xi^{(0)}_{\alpha} (z,p,S;n) = - \tilde \Xi^{(0)\alpha} (z, \tilde p, -\tilde S; \tilde n),\label{eq:Parity2}
\end{align}
where the tilded vector denotes $\tilde p^\mu = (p^0,-\vec p)$.
We emphasize that $S$ in the argument in general specifies the spin state of the hadron $h$. 
In the case of spin-$1/2$ hadron, $S$ is just the polarization vector as usually used and 
we have Eqs.~(\ref{eq:Parity1}) and (\ref{eq:Parity2}) for parity invariance. 
For spin-$1$ hadron, the whole set of variables needed to describe the spin state are more complicated. 
This will be discussed in a precise manner in next section where hadrons with different spins are considered separately. 
It should also be mentioned that, 
as it is known in literature (see e.g. \cite{Collins:1992kk, Jaffe:1993xb,Ji:1993vw}  and the references given there.),
that time-reversal invariance does not constrain the form of fragmentation functions 
because of the presence of final state interactions between the hadron $h$ and remaining multi-hadron state $X$ in the jet. 

Based on the Lorentz covariance, we can analyze the Lorentz structure of these matrix elements 
$\Xi^{(0)}_\alpha$, $\tilde\Xi^{(0)}_\alpha$, $\Xi^{(1)}_{\rho\alpha}$
and $\tilde\Xi^{(1)}_{\rho\alpha}$ in terms of the involved four vectors $p$, $n$ and $S$.  
We express these matrix elements in terms of different Lorentz covariants constructed from $p$, $n$ and $S$ and scalar functions of $z$. 
These functions of $z$ are just different components of fragmentation functions. 
For example, an analysis of $\Xi^{(0)}_\alpha$and $\tilde\Xi^{(0)}_\alpha$ up to twist-3 
is given in \cite{Ji:1993vw}.  
Insert such expressions into Eqs.~(\ref{ltht})-(\ref{t3rchd}), 
we can calculate the hadronic tensors and the differential cross sections 
and obtain their relationship to the fragmentation functions. 
Such relationships are in general different for hadrons with different spins. 
We calculate them for spin-$0$, spin-$1/2$ and spin-$1$ hadrons respectively in next sections. 

We note that calculations of differential cross sections including twist-3 contributions have been 
carried out in liturature such as \cite{Boer:1997mf} for even more complicated process 
$e^++e^-\to h_1+h_2+X$ via electromagnetic interaction for spin-$1/2$ hadrons. 
As most of the higher twist calculations carried out earlier, the approachs 
given there are different from that presented in the current paper in the following way. 
In contrast to what we do here, the calculations given there do not start with a formalism 
after the collinear expansion. 
In stead, they usually start from the hadronic tensors such as 
the $W_{\mu\nu}^{(0)}$, $W_{\mu\nu}^{(1,L)}$ and $W_{\mu\nu}^{(1,R)}$ given by 
Eqs.(\ref{eq:W0}) and (\ref{eq:W1}) obtained directly from the Feynman diagrams 
as given in Figs.~\ref{fdmg}(a) and (b),  
extract the twist-2 and -3 terms by making appropriate approximations during the calculations, 
and insert the gauge link(s) whenever needed to guarantee the gauge invariance. 
It is not studied whether collinear expansion can be applied to such process in a systematic way.   
A systematic formalism is lacking and it is in particular not obvious where the gauge link comes 
from and whether the calculations extend to even higher twists. 
In this way, one usually obatins the same results for leading twist contributions 
as we do using the formalism after the collinear expansion but might get different expressions 
at higher twists since the higher twist correlators are usually different. 
The higher twist correlators used in the formalism after collinear expansion are 
the $\hat\Xi$'s such as those given by Eqs. (\ref{eq:Xi1L}) and (\ref{eq:Xi1R}) 
where covariant derivatives are used while those before the expansion are 
the $\hat\Pi$'s such as those given by Eqs. (\ref{eq:Pi1L}) and (\ref{eq:Pi1R}) 
where gluon fields are used in the corresponding places.  
 
\section{The hadronic tensor up to twist-3}

In order to obtain the hadronic tensor and the cross section,
 we need to expand the 
$\Xi_\alpha^{(0)}$, $\tilde \Xi_\alpha^{(0)}$, $\Xi_{\rho\alpha}^{(1)}$, $\tilde\Xi_{\rho\alpha}^{(1)}$, $\dots$,
according to the Lorentz structure. 
This expansion depends strongly on the spin of the produced hadron $h$. 
In this section, we present the results for the hadronic tensors up to twist-$3$ 
for spin-$0$, spin-$1/2$ and spin-$1$ hadrons respectively. 
The results for the differential cross sections for hadrons with different spins are presented in next section.

\subsection{Spin-0 hadrons}

The situation is simplest for spin zero hadrons such as mesons in the $J^P=0^-$ octet. 
This is also the same if we do not consider the polarization for hadrons with non-zero spins. 

For spin-$0$ hadrons, to the leading twist, we need only to consider, 
\begin{equation}
  z \Xi^{(0)\alpha}  (z,p;n)= p^\alpha D_1(z)  + \cdots  . \label{spin0lt} \\
\end{equation}
By inserting Eq.~(\ref{spin0lt}) into Eq.~(\ref{ltht}), we obtain,
\begin{equation}
\tilde{W}_{\mu\nu}^{(0)}(q,p) =  \frac{1}{2z_B} 
  \mathrm{Tr}\bigl(\hat h^{(0)}_{\mu\nu} {\slashed p}\bigr) D_1(z_B).
\end{equation}
We carry out the trace 
$\mathrm{Tr}(\hat h^{(0)}_{\mu\nu}{\slashed p})=-4( c_1^q d_{\mu\nu}+ i c_3^q \varepsilon_{\perp\mu\nu}) $, 
where $d_{\mu\nu}=g_{\mu\nu}-n_\mu \bar n_\nu -\bar n_\mu n_\nu$,
$\varepsilon_{\perp\mu\nu}=\varepsilon_{\mu\nu\rho\sigma} \bar n^\rho n^\sigma $, $c_1^q = (c_V^{q})^2+(c_A^{q})^2$ and $c_3^q = 2 c_V^q c_A^q$, and obtain,
\begin{equation}
\tilde{W}_{\mu\nu}^{(0)}(q,p) = - \frac{2}{z_B}\left( c_1^q d_{\mu\nu} + i c_3^q \varepsilon_{\perp\mu\nu} \right) D_1(z_B), \label{scalorltwt}
\end{equation}
where $D_1(z)$ is given by,
\begin{equation}
D_1(z) = \frac{z}{4} \sum_X \int \frac{d \xi^-}{2\pi} e^{-ip^+\xi^-/z} \mathrm{Tr} \left[\gamma^+
\langle 0 |  \mathcal{L}^\dagger(0,\infty) \psi(0) |hX\rangle 
\langle hX |\bar\psi(\xi^-) \mathcal{L}(\xi^-,\infty) |0\rangle \right], \label{ffd1}
\end{equation}
is the leading twist fragmentation function in the unpolarized case. 
It can easily be seen that $q^\mu d_{\mu\nu}=0$ and $q^\mu \varepsilon_{\perp\mu\nu} = 0$, 
so that $q^\mu W_{\mu\nu}^{(0)}(q,p)=0$.

For $e^+e^-\to\gamma^*\to q\bar q\to h+X$, i.e., $e^+e^-$ annihilation via electromagnetic 
interaction, the corresponding results can be obtained by putting $c_V^q = 1$ and $c_A^q=0$, i.e., $c_1^q =1$ and $c_3^q=0$
into the above mentioned equations. 
Hence, we have, 
\begin{align}
\tilde W_{\mu\nu}^{(0){\rm em}}(q,p) = - \frac{2}{z_B} d_{\mu\nu} D_1(z_B).
\end{align}

We see that, for the weak interaction, the hadronic tensor given by Eq.~(\ref{scalorltwt}) contains a symmetric and an anti-symmetric part 
while for electromagnetic interaction only the symmetric part left. 

In can easily be shown that the twist-3 contribution in this case is equal to zero.  
This can be shown by analysing the Lorentz structure of 
$\Xi_{\rho\alpha}^{(1)} $ and $\tilde \Xi_{\rho\alpha}^{(1)}$. 
Hence the results given in Eq.~(\ref{scalorltwt}) is also the complete hadronic tensor for spin-0 hadron production up to twist-3. 

\subsection{Spin-1/2 hadrons}

For spin-$1/2$ hadrons, the polarization is described by the polarization vector $S^\mu$. 
At high energies, this polarization vector $S^\mu$ is usually decomposed into 
the transverse polarization vector $S_\perp^\mu$ and the helicity $\lambda_h$ components,
\begin{align}
S^\mu = \lambda_h \frac{p^+}{M} \bar n^\mu + S_\perp^\mu - \lambda_h \frac{M}{2p^+}  n^\mu. \label{eq:polVec}
\end{align}
We see that, compared to the $\bar n$-component,  the $n_\perp$- and the $n$-components are suppressed
by a factor $M/p^+$ and $(M/p^+)^2$ respectively after the Lorentz boost hence contribute only at higher twists.
Up to twist-3,  we need to consider $\tilde{W}_{\mu\nu}^{(0)}(q,p)$ and $\tilde{W}_{\mu\nu}^{(1)}(q,p)$. 
After an analysis of the Lorentz structure of the corresponding correlators 
$\Xi^{(0)\alpha}(z,p,S;n)$, $\tilde\Xi^{(0)\alpha}  (z,p,S;n)$, 
$\Xi^{(1)\rho\alpha} (z,p,S;n)$ and $\tilde \Xi^{(1)\rho\alpha} (z,p,S;n)$ as functions of $p$, $n$ and $S$,  
we obtain that, up to twist-3, we need to consider the following terms, 
\begin{align}
& z \Xi^{(0)\alpha}  (z,p,S;n)=  
p^\alpha D_1(z) +  M \varepsilon_\perp^{\alpha\gamma} S_{\perp\gamma} D_T(z) +\cdots ,\\
& z \tilde\Xi^{(0)\alpha}  (z,p,S;n)=  
\lambda_h p^\alpha \Delta D_{1L}(z)+M S_\perp^\alpha \Delta D_T(z) + \cdots,\\
& z \Xi^{(1)\rho\alpha} (z,p,S;n) =  M \varepsilon_\perp^{\rho\gamma} S_{\perp\gamma} p^\alpha \xi_{\perp S}^{(1)}(z) +\cdots,\\
& z \tilde \Xi^{(1)\rho\alpha} (z,p,S;n) = i M S_\perp^\rho p^\alpha \tilde\xi_{\perp S}^{(1)}(z) +\cdots ,
\end{align}
where all $D_1$, $D_T$, $\Delta D_T$, $\Delta D_{1L}$, $\xi_{\perp S}^{(1)}$, and $\tilde \xi_{\perp S}^{(1)}$ are scalar functions of $z$.

Carrying out the traces, such as, 
$\mathrm{Tr}[h^{(0)}_{\mu\nu}\gamma_\alpha\varepsilon_\perp^{\alpha \gamma}]S_{\perp\gamma}= 
4 (c_1^q  n_{\{\mu}\varepsilon_{\perp\nu\}\gamma} S_\perp^\gamma 
+ i c_3^q n_{[\mu} S_{\perp\nu]})/p^+$,
$\mathrm{Tr}[h^{(0)}_{\mu\nu} \gamma_5 \slashed p] = 4 (c_3^q d_{\mu\nu}+ 
ic_1^q \varepsilon_{\perp\mu\nu})$, 
where $A_{\{\mu}B_{\nu\}}\equiv A_\mu B_\nu+A_\nu B_\mu$ and $A_{[\mu}B_{\nu]}\equiv A_\mu B_\nu-A_\nu B_\mu$,
we obtain that, for spin $1/2$ hadrons, up to twist-3, the hadronic tensors are given by,
\begin{align}
\tilde{W}_{\mu\nu}^{(0)}(q,p,S) = 
\frac{2}{z_B}\Big\{&
-\left(c_1^q d_{\mu\nu}+ i c_3^q \varepsilon_{\perp\mu\nu} \right) D_1(z_B) 
+  \lambda_h\left(  c_3^q d_{\mu\nu}+ i c_1^q \varepsilon_{\perp\mu\nu} \right) \Delta D_{1L}(z_B) \nonumber\\
& + \frac{M}{ p^+} \left( c_1^q n_{\{\mu}\varepsilon_{\perp\nu\}\gamma} S_\perp^\gamma  + i c_3^q n_{[\mu} S_{\perp\nu]}\right) D_T (z_B) \nonumber\\
&-  \frac{M}{ p^+} \left( c_3^q n_{\{\mu} S_{\perp\nu\}}
-i c_1^q n_{[\mu}\varepsilon_{\perp\nu]\gamma} S_\perp^\gamma\right) \Delta D_T (z_B) \Big\},\\
\tilde{W}_{\mu\nu}^{(1,L)}(q,p,S) = &
 \frac{2}{z_B}\frac{M}{p\cdot q} \left( c_1^q p_\nu \varepsilon_{\perp\mu\gamma} S_\perp^\gamma   
- i c_3^q p_\nu S_{\perp\mu}  \right) \big[\xi_{\perp S}^{(1)}(z_B)+\tilde\xi_{\perp S}^{(1)}(z_B)\big]. \label{w1lht}
\end{align}
We see that, besides $D_1(z)$ defined in previous sub-section, there is another leading twist fragmentation function $\Delta D_{1L}(z)$ that contributes to the hadronic tensor and $\Delta D_{1L}(z)$ is defined by,
\begin{align}
&\lambda_h\Delta D_{1L}(z) = \frac{z}{4}  \sum_X \int \frac{d\xi^-}{2\pi} e^{-ip^+\xi^-/z} {\rm Tr}[\gamma^+\gamma_5\langle 0 |   \mathcal{L}^\dagger (0,\infty) \psi(0) |hX\rangle 
\langle hX| \bar\psi(\xi^-) \mathcal{L} (\xi^-,\infty) |0\rangle].
\end{align}
 The two twist-3 fragmentation functions, $D_T(z)$ and $\Delta D_T(z)$, are given by,
\begin{align}
& M S_\perp^2 D_T (z) =  \frac{z}{4}   \sum_X\int \frac{p^+d\xi^-}{2\pi} e^{-ip^+\xi^-/z} \varepsilon_\perp^{\alpha\gamma} S_{\perp\gamma}{\rm Tr}[
\gamma_\alpha\langle 0 |  \mathcal{L}^\dagger (0,\infty) \psi(0) |hX\rangle \langle hX| \bar\psi(\xi^-) \mathcal{L}  (\xi^-,\infty)|0\rangle], \label{t3ffdt}\\
& M S_\perp^2 \Delta D_T  (z)   = \frac{z}{4}   \sum_X\int \frac{p^+d\xi^-}{2\pi} e^{-ip^+\xi^-/z} {\rm Tr}[
\slashed S_\perp\gamma_5\langle 0 |  \mathcal{L}^\dagger (0,\infty) \psi(0) |hX\rangle \langle hX| \bar\psi(\xi^-) \mathcal{L}  (\xi^-,\infty)|0\rangle],
\end{align}
where $S_\perp^2=-|\vec S_\perp|^2$. 
The other two twist-3 fragmentation functions, $\xi_{\perp S}^{(1)}(z)$ and $\tilde\xi_{\perp S}^{(1)}(z)$ are not independent. 
They are related to $D_T(z)$ and $\Delta D_T(z)$.
In fact, using the QCD equation of motion $\gamma\cdot D(x)\psi(x)=0$, we obtain, 
\begin{align}
&\frac{p^+}{z} \Xi^{(0)\rho}(z,p,S;n)
= - n_\alpha \left[ \mathrm{Re} \Xi^{(1)\rho\alpha}(z,p,S;n) + {\varepsilon_\perp^{\rho}}_\sigma \mathrm{Im} \tilde \Xi^{(1)\sigma\alpha}(z,p,S;n) \right],\\
&\frac{p^+}{z} \tilde \Xi^{(0)\rho}(z,p,S;n)
= - n_\alpha \left[ \mathrm{Re} \tilde \Xi^{(1)\rho\alpha}(z,p,S;n) + {\varepsilon_\perp^{\rho}}_\sigma \mathrm{Im} \Xi^{(1)\sigma\alpha}(z,p,S;n)\right] .
\end{align}
This leads to,
\begin{align}
 &D_T (z) = -z \mathrm{Re}[\xi_{\perp S}^{(1)}(z)+ \tilde\xi_{\perp S}^{(1)}(z)], \label{eq:xitoD1}\\
 &\Delta D_T (z) = z \mathrm{Im}[\xi_{\perp S}^{(1)}(z)+\tilde\xi_{\perp S}^{(1)}(z)].\label{eq:xitoD2}
\end{align}
By inserting Eqs.~(\ref{eq:xitoD1}) and (\ref{eq:xitoD2}) into Eq.~(\ref{w1lht}), 
adding the contribution from $\tilde W_{\mu\nu}^{(0)}$ and that from $\tilde W_{\mu\nu}^{(1)}$ together, 
we obtain the complete result for the hadronic tensor up to twist-3 as given by,
\begin{align}
W_{\mu\nu}(q,p,S) =\frac{2}{z_B}\Big\{
 - & \left(c_1^q d_{\mu\nu}+ i c_3^q \varepsilon_\perp^{\mu\nu}  \right) D_1(z_B) 
+ \lambda_h\left( c_3^q d_{\mu\nu}+i c_1^q \varepsilon_{\perp\mu\nu} \right) \Delta D_{1L}(z_B)  \nonumber\\
+ &\frac{M}{p\cdot q}  \left[ c_1^q (q-2p/z_B)_{\{\mu}\varepsilon_{\perp\nu\}\gamma} S_\perp^\gamma 
+ i c_3^q (q-2p/z_B)_{[\mu} S_{\perp\nu]}  \right] D_T (z_B) \nonumber\\
- &\frac{M}{p\cdot q}  \left[  c_3^q (q-2p/z_B)_{\{\mu} S_{\perp\nu\}}
- i c_1^q (q-2p/z_B)_{[\mu}\varepsilon_{\perp\nu]\gamma} S_\perp^\gamma  \right] \Delta D_T (z_B)\Big\}. \label{fermiontwist3}
\end{align}
It is easy to verify that $q^\mu W_{\mu\nu}=0$.

We note that the first term in this case is the same as that obtained in the case for spin zero hadrons. 
The other terms are spin dependent hence do not exist for the spin zero case. The second term depends on the longitudinal 
component of the polarization while the other terms depend on the transverse components of the polarization. 
We will come back to this point in the next section. 

If we consider $e^+e^-\to\gamma^*\to q\bar q\to h+X$,  we have,
\begin{align}
W_{\mu\nu}^{\rm em}(q,p,S) =\frac{2}{z_B}\Big\{
 - &  d_{\mu\nu} D_1(z_B) 
+ i\lambda_h  \varepsilon_{\perp\mu\nu}  \Delta D_{1L}(z_B)  \nonumber\\
+ & \frac{M}{p\cdot q}   (q-2p/z_B)_{\{\mu}\varepsilon_{\perp\nu\}\gamma} S_{\perp}^\gamma 
 D_T (z_B) \nonumber\\
+& i\frac{M}{p\cdot q}   
  (q-2p/z_B)_{[\mu}\varepsilon_{\perp\nu]\gamma} S_\perp^\gamma  \Delta D_T (z_B)\Big\}.
\end{align}
We see that the terms remained include a symmetric spin independent leading twist term, an anti-symmetric longitudinal spin dependent leading twist term and also a twist-3 transverse spin dependent term.  They can give us measurable effects that we will discuss in next section.

\subsection{Vector mesons}

For spin one particle, we need to use the $3\times 3$ spin density matrix $\rho$ to describe the polarization state. 
We all know that $\rho$ is a Hermite and normalized (i.e. $\rm{Tr} \rho=1$) matrix hence has 8 degrees of freedom. 
This means that we need 8 independent variables to describe the polarization state of the vector meson. 
We choose to decompose $\rho$ into a polarization vector $S^\mu$ and a polarization tensor $T^{\mu\nu}$. 
In the rest frame of the vector meson, $\rho$ takes the following form~\cite{Bacchetta:2000jk}, 
\begin{equation}
\rho=\frac{1}{3}\left(1+\frac{3}{2}S^i\Sigma^i+3T^{ij}\Sigma^{ij}\right),
\end{equation}
where $\Sigma^i$ is the spin matrix for spin one state, $\Sigma^{ij}$ is defined as,
\begin{equation}
\Sigma^{ij}=\frac{1}{2}\left( (\Sigma^i\Sigma^j+\Sigma^j\Sigma^i)-\frac{2}{3}\delta_{ij}\right) .
\end{equation}
$T^{ij}$ is a traceless symmetric tensor and is parameterized in terms of $S_{LL}$, $S_{LT}^i$ and $S_{TT}^{ij}$,
\begin{equation}
\mathbf{T}= \frac{1}{2}
\left(
\begin{array}{ccc}
-\frac{2}{3}S_{LL} + S_{TT}^{xx} & S_{TT}^{xy} & S_{LT}^x  \\
S_{TT}^{xy}  & -\frac{2}{3} S_{LL} - S_{TT}^{xx} & S_{LT}^{y} \\
S_{LT}^x & S_{LT}^{y} & \frac{4}{3} S_{LL}
\end{array}
\right),
\end{equation}
so that the spin density matrix $\rho$ is given by, 
\begin{equation}
\rho= 
\left(
\begin{array}{ccc} 
\frac{1 + S_{LL}}{3} + \frac{S_L}{2} &  \frac{(S_{LT}^x-iS_{LT}^y)+ (S_T^x - i S_T^y)}{2\sqrt{2}} & \frac{S_{TT}^{xx}-iS_{TT}^{xy}}{2} \\
\frac{(S_{LT}^x + iS_{LT}^y) + (S_T^x + i S_T^y)}{2\sqrt{2}}  & \frac{1-2S_{LL}}{3} &  \frac{(-S_{LT}^{x}+iS_{LT}^y)+(S_T^x-iS_T^y)}{2\sqrt{2}}\\
\frac{S_{TT}^{xx}+iS_{TT}^{xy}}{2} & \frac{(-S_{LT}^{x}-iS_{LT}^y)+(S_T^x+iS_T^y)}{2\sqrt{2}} & \frac{1 + S_{LL}}{3} - \frac{S_L}{2}
\end{array}
\right), \label{eq:rhoV}
\end{equation}

The polarization vector $S^\mu$ is similar to what we have for spin-$1/2$ particle and in  
a moving frame of the vector meson, $S^\mu$ behaves as a Lorentz vector in the same form as 
that given in Eq.~(\ref{eq:polVec}) and satisfies $p\cdot S=0$.  
The physical meaning of the polarization vector $S^\mu$ is also clear and is similar to that for spin-$1/2$ particle. 
$T^{\mu\nu}=T^{\nu\mu}$ is a symmetric Lorentz tensor satisfying $p_\mu T^{\mu\nu}=0$.
The different components of $T^{\mu\nu}$ have also clear physical significances. 
The ranges of values of these parameters are e.g. 
$-1 \leq S_{LL} \leq \frac{1}{2} $,  $-1 \leq S_{LT}^i \leq 1$,  and $-1 \leq S_{TT}^{ij} \leq 1$. 
From Eq.~(\ref{eq:rhoV}), we see clearly that $S_{LL}$ is directly related to $\rho_{00}$ that describes the so-called spin alignment of vector meson. 
 Other components of $T^{ij}$ are related to the probabilities for the vector meson to be in 
 different transversely polarized states. 
 A detailed description can e.g. be found in the appendix of~\cite{Bacchetta:2000jk}.

Using such a decomposition of $\rho$, we should obtain the quark correlators as functions of $n$, $p$,  $S^\mu$, $S_{LL}$, $S_{LT}^{\mu}$ and $S_{TT}^{\mu\nu}$. 
For the inclusive process, the contributing terms up to twist-3 are given by,
\begin{align}
& z \Xi^{(0)\alpha} (z,p,S;n) = p^\alpha \left[ D_1(z) + S_{LL} D_{1LL}(z) \right] +  
  M \varepsilon_\perp^{\alpha\gamma} S_{\perp\gamma} D_T(z) +  M S_{LT}^\alpha D_{LT}(z) + \cdots,\\
& z \tilde \Xi^{(0)\alpha} (z,p,S;n) = \lambda_h p^\alpha \Delta D_{1L}(z) +  M S_\perp^\alpha \Delta D_T(z)  + 
  M \varepsilon_\perp^{\alpha\gamma} S_{LT,\gamma} \Delta D_{LT}(z) +\cdots,\\
& z \Xi^{(1)\rho\alpha} (z,p,S;n) =  p^\alpha\left[M \varepsilon_\perp^{\rho\gamma} S_{\perp\gamma}  \xi_{\perp S}^{(1)}(z)+
  M S_{LT}^\rho  \xi_{LTS}^{(1)}(z)\right]  +\cdots,\\
& z \tilde \Xi^{(1)\rho\alpha} (z,p,S;n) = i p^\alpha \left[M S_\perp^\rho  \tilde \xi_{\perp S}^{(1)} (z) + 
 i  M \varepsilon_\perp^{\rho\gamma} S_{LT,\gamma}   \tilde\xi_{LTS}^{(1)}(z)\right]+ \cdots.
\end{align}

By inserting the above expansion into Eqs.~(\ref{ltht})-(\ref{t3rchd}), carrying out the traces and by making use of the relationships such as,
\begin{align}
D_{LT}(z)& = -z \mathrm{Re}[\xi_{LTS}^{(1)}(z) - \tilde \xi_{LTS}^{(1)}(z)],\\
\Delta D_{LT}(z)& =- z \mathrm{Im}[\xi_{LTS}^{(1)}(z) - \tilde\xi_{LTS}^{(1)}(z)].
\end{align} 
derived from the equation of motion to replace $\xi_{LTS}^{(1)}$ or $\tilde\xi_{LTS}^{(1)}$ by $D_{LT}$ or $\Delta D_{LT}$'s, 
we obtain the hadronic tensor for vector meson as, 
\begin{align}
W_{\mu\nu}(q,p,S) =\frac{2}{z_B}\Big\{
 - & \left(c_1^q d_{\mu\nu}+ i c_3^q \varepsilon_\perp^{\mu\nu}  \right) [ D_1(z_B) + S_{LL} D_{1LL}(z_B)]
+ \lambda_h\left( c_3^q d_{\mu\nu}+i c_1^q \varepsilon_{\perp\mu\nu} \right) \Delta D_{1L}(z_B)  \nonumber\\
+  \frac{M}{p\cdot q} & \left[ c_1^q (q-2p/z_B)_{\{\mu}\varepsilon_{\perp\nu\}\gamma} S_\perp^\gamma 
+ i c_3^q (q-2p/z_B)_{[\mu} S_{\perp\nu]}  \right] D_T (z_B) \nonumber\\
 +\frac{M}{p\cdot q} & \bigl[ c_1^q (q-2p/z_B)_{\{\mu} S_{LT,\nu\}} 
  - i c_3^q (q-2p/z_B)_{[\mu}\varepsilon_{\perp\nu]\gamma} S_{LT}^\gamma\bigr] D_{LT}(z_B) \nonumber\\
-  \frac{M}{p\cdot q} & \left[  c_3^q (q-2p/z_B)_{\{\mu} S_{\perp\nu\}}
- i c_1^q (q-2p/z_B)_{[\mu}\varepsilon_{\perp\nu]\gamma} S_\perp^\gamma  \right] \Delta D_T (z_B)\nonumber\\
-  \frac{M}{p\cdot q} & \bigl[ c_3^q(q-2p/z_B)_{\{\mu}\varepsilon_{\perp\nu\}\gamma} S_{LT}^\gamma 
+i c_1^q (q-2p/z_B)_{[\mu} S_{LT,\nu]} \bigr] \Delta D_{LT}(z_B) \Bigr\}.  \label{hadtenmes}
\end{align}

Again, we can obtain the corresponding hadronic tensor for $e^+e^-\to\gamma^*\to q\bar q\to h+X$ 
by putting $c_1=1$ and $c_3=0$ into Eq.~(\ref{hadtenmes}), and it is given by,
\begin{align}
W_{\mu\nu}^{\rm em}(q,p,S) =\frac{2}{z_B}\Big\{
 - &  d_{\mu\nu} [ D_1(z_B) + S_{LL} D_{1LL}(z_B)]
+ i\lambda_h  \varepsilon_{\perp\mu\nu}  \Delta D_{1L}(z_B)  \nonumber\\
+ & \frac{M}{p\cdot q}   (q-2p/z_B)_{\{\mu}\varepsilon_{\perp\nu\}\gamma} S_\perp^\gamma 
 D_T (z_B) \nonumber\\
+ & \frac{M}{p\cdot q}  (q-2p/z_B)_{\{\mu} S_{LT,\nu\}} 
   D_{LT}(z_B) \nonumber\\
+ & i \frac{M}{p\cdot q} 
   (q-2p/z_B)_{[\mu}\varepsilon_{\perp\nu]\gamma} S_\perp^\gamma   \Delta D_T (z_B)\nonumber\\
- & i \frac{M}{p\cdot q}  (q-2p/z_B)_{[\mu} S_{LT,\nu]}  \Delta D_{LT}(z_B) \Bigr\}.  \label{hadtenmes-em}
\end{align}

From the results given by Eqs.~(\ref{hadtenmes}) and (\ref{hadtenmes-em}), we see similar structure as that in the case for spin-$1/2$ hadrons, i.e., a spin independent leading twist term that is the same as in the case for spin zero hadrons, 
a longitudinal polarization dependent leading twist term, and a number of transverse spin dependent twist-3 terms. 
We have, for vector mesons, in particular also a leading twist $S_{LL}$ term which is related to the spin alignment and 
we will discuss in detail in next section.

\section{The cross section and polarization of hadrons produced}

By inserting the hadronic tensors obtained in last section into Eq.~(\ref{CS}), we obtain the differential 
cross sections in the corresponding cases.  
From the cross sections, we obtain not only the production rates of the hadrons but also 
the polarization of the hadrons produced in different cases. 
In this section, we present the results for hadrons with different spins respectively. 
In this paper, we consider only the reactions with unpolarized electrons and unpolarized positions. 

\subsection{Spin-0 hadrons}

By inserting Eq.~(\ref{scalorltwt}) into Eq.~(\ref{CS}), we obtained the differential cross section for inclusive hadron production in $e^+e^-$ annihilation as, 
\begin{equation}
E_p\frac{d\sigma}{d^3 p} = \frac{2\alpha^2}{z Q^4} \chi T_0(y) D_1(z), \label{Z1}
\end{equation}
where $\alpha=e^2/4\pi$ is the fine structure costant, 
$\chi={Q^4}/{[(Q^2-M_Z^2)^2+\Gamma_Z^2 M_Z^2]\sin^42\theta_W}$ is a kinematic factor depending on $Z^0$ mass and Weinberg angle,
the coefficient $T_0$ is a function of $y$ and is given by,
\begin{align}
T_0^q(y) =   c_1^q c_1^e A(y)  - c_3^q c_3^e B(y) , \label{c0}
\end{align}
and $A(y)=(1-y)^2+y^2, B(y)=1-2y$.
Here, $y$ is the longitudinal momentum fraction of electron defined as, 
$y\equiv l_1\cdot n / k \cdot n =zl_1^+/p^+$ so that 
$l_1 = y p^+ \bar n/z + (1-y) z Q^2 n/(2p^+) + l_\perp$, 
$l_\perp=(0,0,l_{\perp x},0)$, $|\vec l_\perp|=|l_{\perp x}|=\sqrt{y(1-y)}Q$. 
In the $e^+e^-$ center of mass frame, $y = (1+ \cos \theta) /2$ 
where $\theta$ is the angle between the incident electron and the produced quark.
In terms of $\theta$, $A(y)=(1+\cos^2\theta)/2$ and $B(y)=-\cos\theta$.
The coefficient function $T_0(y)$ is flavor dependent and is essentially the relative weight 
for the contribution from the given flavor.

We note that the differential cross section is in general a function of $z=z_B$ and $y$. 
We can change the variables and obtain the differential cross section with respect to $z$ and $y$ as,
\begin{equation}
\frac{d^2\sigma}{dzdy}= \frac{2\pi\alpha^2}{Q^2} \chi  T_0(y) D_1(z) .
\end{equation}
We emphasize once more that $z=p^+/k^+$ is the light cone momentum fraction of the quark carried 
by the hadron and $y=l_1^+/k^+$ is the light cone momentum fraction of the incident electron which is 
determined by the angle between the electron and the quark and is given by  
$y=\cos^2 (\theta/2)$ in the c.m. frame of $e^+e^-$.     
The $y$ or $\theta$ dependence is contained in the coefficient function $T_0(y)$.
We can carry out the integration over $y$ or $\theta$ and obtain,
\begin{equation}
\frac{d\sigma}{dz} = \frac{2\pi\alpha^2}{Q^2} \chi t_0  D_1(z) ,
\end{equation}
where $t_0=\int dy T_0(y)= 2c_1^q c_1^e /3$ is a flavor dependent constant.  

The corresponding results for $e^+e^- \to \gamma^* \to q\bar q\to h+X$ are obtained by putting 
$c_1^q=1$ and $c_3^q=0$ into the corresponding equations. 
In this case, we have, $T_0(y)= A(y)$, independent of the flavor, 
the kinematic factor $\chi=1$ and $g_z^4/16$ should be replaced by $e^4 e_q^2$.
Hence, 
\begin{equation}
E_p\frac{d\sigma^{\rm em}}{d^3 p} = \frac{2\alpha^2 e_q^2}{ z Q^4} [(1-y)^2+y^2] D_1(z).
\end{equation}
In terms of $z$ and $y$, we have,
\begin{equation}
\frac{d\sigma^{\rm em}}{dzdy}= \frac{2\pi \alpha^2 e_q^2}{ Q^2}  [ (1-y)^2 + y^2 ] D_1(z) .
\end{equation}
Carrying out the integration over $y$ or $\theta$, we have,
\begin{equation}
\frac{d\sigma^{\rm em}}{dz}=  \frac{4\pi \alpha^2e_q^2}{3 Q^2}  D_1(z) .
\end{equation}

If we write out the summations over flavor and color explicitly, we have, e.g., 
\begin{align}
\frac{d^2\sigma}{dzdy}= N_c \sum_{q} \frac{2\pi\alpha^2}{Q^2} \chi T_0^{q}(y) D_1^{q\to h}(z) ,
\end{align}
where the sum over $q$ runs for all quark and anti-quark flavors involved, and for anti-quark, 
 it can easily be seen that $T_0^{\bar q}(y)= T_0^q (1-y)$,
 and the fragmentation function is defined as,
\begin{align}
D_1^{\bar q\to h}(z) = \frac{z}{4} \sum_X\int \frac{d\xi^-}{2\pi}e^{-ip^+\xi^-/z}
   \langle 0 | \bar\psi(0)\gamma^+ \mathcal{L} (0,\infty) |hX \rangle \langle hX| \mathcal{L}^\dagger (\xi^-,\infty) \psi(\xi^-,\infty)|0\rangle .
\end{align}

For $e^+e^- \to \gamma^* \to q\bar q\to h+X$, the cross section takes the form, 
\begin{equation}
\frac{d\sigma^{\rm em}}{dz}=  \sum_q \frac{4\pi \alpha^2}{3 Q^2} e_q^2 D_1^{q\to h+X}(z) .
\end{equation}
which is just the result used usually when describing hadron production in $e^+e^-$ annihilation at high energies in the unpolarized case.

\subsection{Spin-$1/2$ hadrons}
For hadrons with nonzero spins, we can calculate not only the differential cross section but also the polarizations. 
Here, we present the results for cross section and polarization for spin-1/2 hadrons.

\subsubsection{The cross section}
We insert the hadronic tensor given by Eq.~(\ref{fermiontwist3}) into Eq.~(\ref{CS}), and we obtain 
the differential cross section for spin-$1/2$ hadrons as,
\begin{align}
E_p \frac{d\sigma}{d^3p} = \frac{2\alpha^2}{zQ^4}\chi\Bigl\{ 
& \bigl[ T_0 (y) D_1 (z)  +  \lambda_h T_1 (y) \Delta D_{1L} (z) \bigr]   \nonumber\\ 
+& \frac{2M}{ p\cdot q} \bigl[  \varepsilon_\perp^{l_\perp S_\perp} T_2 (y) D_T (z) 
+ l_\perp \cdot S_\perp T_3(y) \Delta D_T (z) \bigr]\Bigr\}.
\end{align}
We see that, besides the first term that is equivalent to what we have for spin zero hadrons, 
there are three other spin dependent terms where the coefficient functions $T_i(y)$'s for quarks are given by,
\begin{align}
T_1 (y)&=- c_3^q c_1^e A(y)+ c_1^q c_3^e B(y),\\
T_2 (y)&=- c_3^q c_3^e +  c_1^q c_1^e B(y) ,\\
T_3 (y)&= c_1^q c_3^e -  c_3^q c_1^e B(y),
\end{align}
and these for the anti-quarks are related to those for the corresponding quarks in the following way, 
\begin{align}
T_1^{\bar q} (y)&= T_1^q (1-y),\\
T_2^{\bar q} (y)&= -T_2^q (1-y) ,\\
T_3^{\bar q} (y)&= -T_3^q (1-y).
\end{align}
We see also that $T_2(y)$ and $T_3(y)$ are just the first derivative of $T_0(y)$ and $T_1(y)$ respectively, i.e., 
$T_2(y)=-(1/2)d T_0^q (y)/d y$ and $T_3(y)=-(1/2)d T_1 (y)/d y$.   

Denote the angle between $\vec{S}_\perp$ and $\vec{l}_\perp$ by $\phi_s$, 
we obtain $\varepsilon_\perp^{l_\perp S_\perp}=|l_\perp||S_\perp|\sin \phi_s$ 
and $l_\perp\cdot S_\perp = - |l_\perp||S_\perp| \cos \phi_s$.
$|l_\perp| = \sqrt{y(1-y)}Q = \sin \theta Q/2 $. 
So that the cross section can also be expressed as, 
\begin{align}
E_p \frac{d\sigma}{d^3p} &=\chi\frac{2\alpha^2}{zQ^4}  \Bigl\{ 
 \bigl[ T_0 (y) D_1 (z)  +  \lambda_h T_1 (y) \Delta D_{1L} (z) \bigr]   \nonumber\\ 
 &+ \frac{4M}{z Q}|\vec S_\perp| \sqrt{y(1-y)} 
\bigl[ T_2(y) D_T(z)\sin \phi_s  - T_3(y)\Delta D_T(z) \cos \phi_s \bigr]\Bigr\}.
\end{align}
We see that $\Delta D_{1L} (z)$ is responsible for the longitudinal polarization of the hadron while 
$\Delta D_T(z)$ and $D_T(z)$ are sources of the transverse polarizations in and transverse to the leptonic plane respectively. 
We will come back to this point in next sub-section. 

In terms of $z$ and $y$, we have,
\begin{align}
\frac{d\sigma}{dzdy} = \chi\frac{2\pi\alpha^2}{Q^2}  \Bigl\{ 
& \bigl[ T_0 (y) D_1 (z)  +  \lambda_h T_1 (y) \Delta D_{1L} (z) \bigr]   \nonumber\\ 
+\frac{4M}{z Q}|\vec S_\perp| & 
\bigl[ \tilde T_2(y) D_T(z)\sin \phi_s  - \tilde T_3(y)\Delta D_T(z) \cos \phi_s \bigr]\Bigr\}, \label{fermion}
\end{align}
where $\tilde T_i(y) = \sqrt{y(1-y)} T_i(y)$.
Carrying out the integration over $y$ (or $\theta$), we have,
\begin{align}
\frac{d\sigma}{dz} = \chi\frac{2\pi\alpha^2}{Q^2}  \Bigl\{ 
&  \bigl[ t_0 D_1 (z) + \lambda_h t_1 \Delta D_{1L} (z) \bigr]   \nonumber\\ 
+ \frac{4M}{z Q}|\vec S_\perp| & 
\bigl[  \tilde t_2 D_T(z)\sin \phi_s - \tilde t_3 \Delta D_T(z) \cos \phi_s \bigr]\Bigr\},
\end{align}
where $t_i \equiv \int dy T_i(y)$ are flavor dependent constants determined by $c_1^q$ and $c_3^q$, i.e., 
$t_1 = -2 c_3^q c_1^e / 3 $,
$\tilde t_2 = -\pi c_3^q c_3^e/8 $
and 
$\tilde t_3 = \pi c_1^q c_3^e /8$ .

By inserting $c_1=1$ and $c_3=0$ into these equations, we obtain the corresponding results for $e^+e^-\to\gamma^*\to q\bar q\to h+X$, where we have, $T_1(y) = \tilde T_3(y) = 0$,  
and $\tilde T_2(y)=\sqrt{y(1-y)}B(y) = -\sin 2\theta /2$.
Hence, the cross section is given by,
\begin{align}
E_p \frac{d\sigma^{\rm em}}{d^3p} = \frac{2 \alpha^2 e_q^2}{zQ^4}  \Bigl\{  D_1 (z)  (1+\cos^2\theta) 
- |\vec S_\perp|\frac{4M}{z Q}  D_T (z) \sin 2\theta \sin \phi_s \Bigr\}.
\end{align}
In terms of $z$ and $y$, we have,
\begin{equation}
\frac{d\sigma^{\rm em}}{dzdy}= \frac{2\pi \alpha^2 e_q^2}{ Q^2}  \Bigl\{ [(1-y)^2+y^2] D_1 (z) 
- |\vec S_\perp|\frac{4M}{z Q} \sqrt{y(1-y)} (1-2y)D_T (z) \sin \phi_s \Bigr\}.
\end{equation}
Carrying out the integration over $y$, we see that all the twist-3 terms vanish and 
we obtain, 
\begin{equation}
\frac{d\sigma^{\rm em}}{dz}=  \frac{4\pi \alpha^2}{3 Q^2} e_q^2 D_1 (z) ,
\end{equation}
which is the same as that obtained for the spin-0 hadron. 
  
\subsubsection{Hadron polarization}
From Eq.~(\ref{fermion}), we see that the spin-$1/2$ hadron produced in 
$e^+e^-\to Z\to q\bar q\to h+X$ is longitudinally polarized.   
The longitudinal polarization  is given by,
\begin{equation}
P_{Lh}(z,y)=\frac{T_1(y)\Delta D_{1L}(z)}{T_0(y)D_{1}(z)}, \label{eq:Lpol}
\end{equation}
We write out the flavor index and summation over the flavor explicitly so that Eq.~(\ref{eq:Lpol}) takes 
the following form, 
\begin{equation}
P_{Lh}(z,y)=\frac{\sum_{q} T_{1}^{q} (y)\Delta D_{1L}^{q\to h}(z)}{\sum_{q} T_{0}^{q}(y)D_1^{q\to h}(z)}. \label{polarizationf}
\end{equation}

We recall that $T_{0}^q(y)$ represents the relative weight for the contribution from quark (anti-quark) 
of flavor $q$ and Eq.~(\ref{polarizationf}) can be re-written as, 
\begin{equation}
P_{Lh}(z,y)=
\frac{\sum_{q} P_{q}(y) T_{0}^{q}(y)\Delta D_{1L}^{q\to h}(z)}{\sum_{q} T_{0}^{q}(y)D_1^{q\to h}(z)},
\end{equation}
where $P_{q}(y)=T_{1}^{q}(y)/T_{0}^{q}(y)$ is the polarization of the quark produced. 
Such quark polarization has been calculated explicitly in e.g.~\cite{Augustin:1978wf} 
and the numerical results can be found there.
It is also clear that $\Delta D_{1L}^{q\to h}(z)$ is nothing else but the spin transfer in the 
fragmentation process. 

We see that the polarization is in general different for hadrons produced in different $\theta$ directions. 
The $\theta$ or $y$ dependence comes from the $y$ dependence of $T_i$ which describes 
the relative weights and polarizations of the quarks of different flavors. 
To study the fragmentation functions, we can integrate over $y$ or $\theta$ and obtain,
\begin{equation}
P_{Lh}(z)=
\frac{\sum_{q} t_{0}^{q} P_{q}\Delta D_{1L}^{q\to h}(z)}{\sum_{q} t_{0}^{q} D_1^{q\to h}(z)}, \label{ltlp}
\end{equation}
where $P_{q}=t_{1}^{q} /t_{0}^{q} =- c_3^q / c_1^q $, 
is the polarization of the quark of flavor $q$ averaged over different directions. 

It is also very interesting to see, from Eq.~(\ref{fermion}), that although the quark and/or anti-quark is 
longitudinally polarized in $e^+e^-\to Z\to q\bar q\to h+X$,  the produced hadron $h$ can possess also 
a transverse polarization at the twist-3 level. 
We take the helicity frame of $h$, i.e., take the direction of motion of $h$ as $z$-direction, 
and we obtain, 
\begin{align}
&P_{hx}(z,y)=-\frac{4M}{zQ} \frac{\sum_{q} \tilde T_{3}^{q}(y) \Delta D_T^{q\to h}(z)}{ \sum_{q} T_{0}^{q}(y)D_1^{q\to h}(z)},\\
&P_{hy}(z,y)= \frac{4M}{zQ} \frac{\sum_{q} \tilde T_{2}^{q}(y) D_T^{q\to h}(z)}{\sum_{q} T_{0}^{q} (y)D_1^{q\to h}(z)},
\end{align}
for given $y$ or $\theta$.
Here, we recall once more that the $x$ and $y$ directions are defined in or transverse to the leptonic plane.
Integrating over $y$, we obtain,
\begin{align}
&P_{hx}(z)=-\frac{4M}{zQ} \frac{\sum_{q} \tilde t_{3}^{q} \Delta D_T^{q\to h}(z)}{ \sum_{q} t_{0}^{q} D_1^{q\to h}(z)},\\
&P_{hy}(z)= \frac{4M}{zQ} \frac{\sum_{q} \tilde t_{2}^{q} D_T^{q\to h}(z)}{\sum_{q} t_{0}^{q} D_1^{q\to h}(z)}.
\end{align}

If we consider $e^+e^-\to\gamma^*\to q\bar q\to h+X$, we see that the longitudinal polarization and 
the transverse polarization inside the leptonic plane vanish, i.e., $P_{Lh}^{em} (z,y)= P_{hx}^{em} (z,y)=0$. 
However, we can still have a non-vanishing polarization transverse to the leptonic plan at the twist-3 level. 
The result is given by, 
\begin{align}
&P_{hy}^{\rm em} (z,y)=\frac{4M}{zQ} \frac{\sqrt{y(1-y)}(1-2y)}{(1-y)^2+y^2} \frac{\sum_{q} e_q^2D_T^{q\to h}(z) }{\sum_{q} e_q^2D_1^{q\to h} (z)},
\end{align}
or in terms of the angle $\theta$, 
\begin{align}
&P_{hy}^{\rm em} (z,\theta)=-\frac{2M}{zQ} \frac{\sin2\theta}{1+\cos^2\theta} \frac{\sum_{q} e_q^2D_T^{q\to h}(z) }{\sum_{q} e_q^2D_1^{q\to h} (z)}.
\end{align}
This polarization vanishes also after the integration over $y$ or $\theta$, i.e.,  $P_{Lh}^{em} = P_{hx}^{em} = P_{hy}^{em} = 0$.

We note that such a transverse polarization has also been expected in \cite{Boer:1997mf}  
where calculations of differential cross section of two hadron production $e^+e^-\to h_1+h_2+X$ 
have been carried out starting directly from the hadronic tensor reading from the diagrams 
similar to those given by Fig. \ref{fdmg}.  
The results take the same form when appropriate gauge link is inserted 
the fragmentation functions given there.  

Experimental studies on the longitudinal polarization of $\Lambda$-hyperon have been carried out 
by ALEPH and OPAL Collaborations at LEP~\cite{Buskulic:1996vb, Ackerstaff:1997nh}. 
The data show a clear polarization and can be used to study the properties in general and  to obtain 
a parameterization of $\Delta D_{1L}(z)$ in particular. 
Such parameterizations exist already in literature and can be found e.g. in~\cite{deFlorian:1997zj}. 
We will not go to the details in that direction in this paper.

Little discussion can be found on the transverse polarization presented above for $e^+e^-$ annihilation and there is no measurement available yet.
We emphasize that such measurements are very useful in studying higher effects in general and provide us direct information on the twist-3 fragmentation function given in Eq.~(\ref{t3ffdt}) in particular. 

\subsection{Vector meson}
For hadrons with spin-one e.g. the vector mesons, the spin dependence is more complicated thus makes 
the study even more interesting. Here, we present the results for the differential cross section and the results 
for the spin alignment factor $\rho_{00}$ in the following.
   
\subsubsection{The cross section}
By insert the hadronic tensor Eq.~(\ref{hadtenmes}) into Eq.~(\ref{CS}), we get the cross section,
\begin{align}
E\frac{d\sigma}{d^3p} =  \frac{2\alpha^2}{zQ^4}  \chi
   & \Bigl\{ \big[ T_0 (y) D_1(z) +  T_0(y) S_{LL}  D_{1LL}(z_B) + \lambda_h T_1 (y) \Delta D_{1L}(z)   \bigr] \nonumber\\
+ \frac{4M}{zQ}  |\vec S_\perp| &  \bigl[ \tilde T_2(y) \sin \phi_s D_T(z) - \tilde T_3 (y)  \cos \phi_s \Delta D_T(z) \bigr]   \nonumber\\
+ \frac{4M}{zQ}  |\vec S_\perp| & \bigl[  - \tilde T_2 (y)  \cos \phi_{LT} D_{LT}(z)  + \tilde T_3 (y) \sin \phi_{LT} \Delta D_{LT}(z) \bigr] \Bigr\},
\end{align}
where $\phi_{LT}$ is the angle between $\vec{S}_{LT}$ and $\vec{l}_\perp$.
We see that the cross section in general depends on the polarization of the vector meson. 
We also see that the coefficient functions $T_i(y)$ describe the relative weights and polarizations of 
the quarks and/or anti-quarks of different flavors. They are the same as those defined in Sec. IVB
for production of spin-1/2 hadrons.

In terms of $z$ and $y$, we have,
\begin{align}
\frac{d\sigma}{dzdy} =  \chi\frac{2\pi\alpha^2}{Q^2}
& \Bigl\{ \big[ T_0 (y) D_1(z) +  T_0(y) S_{LL}  D_{1LL}(z_B) + \lambda_h T_1 (y) \Delta D_{1L}(z)   \bigr] \nonumber\\
+ \frac{4M}{zQ}  |\vec S_\perp| & \bigl[ \tilde T_2(y) \sin \phi_s D_T(z) - \tilde T_3 (y)  \cos \phi_s \Delta D_T(z) \bigr]   \nonumber\\
+ \frac{4M}{zQ}  |\vec S_\perp| & \bigl[ - \tilde T_2 (y)  \cos \phi_{LT} D_{LT}(z) + \tilde T_3 (y) \sin \phi_{LT} \Delta D_{LT}(z) \bigr] \Bigr\}. \label{vectormeson}
\end{align}
Carrying out the integration over $y$ or $\theta$, we have,
\begin{align}
\frac{d\sigma}{dz} = \frac{2\pi\alpha^2}{Q^2}\chi
& \Bigl\{ \big[ t_0 D_1(z) +  t_0 S_{LL}  D_{1LL}(z) + \lambda_h t_1 \Delta D_{1L}(z)   \bigr] \nonumber\\
+ \frac{4M}{zQ}  |\vec S_\perp| &  \bigl[ \tilde t_2 \sin \phi_s D_T(z) - \tilde t_3 \cos \phi_s \Delta D_T(z) \bigr]   \nonumber\\
+ \frac{4M}{zQ}  |\vec S_\perp| &  \bigl[ - \bar t_2 \cos \phi_{LT} D_{LT}(z) + \tilde t_3 \sin \phi_{LT} \Delta D_{LT}(z) \bigr] \Bigr\}. \label{vectormeson2}
\end{align}

For the electromagnetic interaction process $e^+e^-\to\gamma^*\to q\bar q\to h+X$, the corresponding result is obtained by putting 
$T_0(y)=A(y)$, $T_1(y)=\tilde T_3(y)=0$, $\tilde T_2(y)= \sqrt{y(1-y)}B(y)$, and we have, 
\begin{align}
E\frac{d\sigma^{\rm em}}{d^3p} = \frac{2 \alpha^2 e_q^2 }{ Q^4 z}  
   & \Bigl\{ A (y) \big[ D_1(z) +  S_{LL}  D_{1LL}(z)  \bigr] \nonumber\\
+ & \frac{M}{zQ^2}  B(y) \bigl[   \varepsilon_\perp^{l_\perp S_\perp} D_T(z)  + l_\perp \cdot S_{LT} D_{LT}(z)     \bigr] \Big\}.
\end{align}
In terms of $z$ and $y$, we have,
\begin{align}
\frac{d\sigma^{\rm em}}{dzdy}= \frac{2\pi  \alpha^2 e_q^2 }{Q^2 }  
& \Bigl\{ A (y) \bigl[ D_1(z) +  S_{LL}  D_{1LL}(z)  \bigr] \nonumber\\
+ \frac{M}{zQ}  & \sqrt{y(1-y)} B(y) \bigl[  |\vec S_\perp| \sin \phi_s D_T(z)  - |\vec S_{LT}| \cos \phi_{LT} D_{LT}(z)     \bigr] \Big\}.
\end{align}
Carrying out the integration over $y$, we obtain, 
\begin{equation}
\frac{d\sigma^{\rm em}}{dz}= e_q^2 \frac{4\pi \alpha^2}{3 Q^2 } \left[ D_1(z) + S_{LL} D_{1LL}(z) \right].
\end{equation}

\subsubsection{The spin alignment}

Polarization of vector meson  has been studied in~\cite{Ackerstaff:1997kd, Ackerstaff:1997kj, Abreu:1997wd} 
by OPAL and DELPHI at LEP where $\rho_{00}$ has been measured in the helicity frame of the vector meson. 
Phenomenological studies have also been carried out in e.g.~\cite{Xu:2001hz}. 
From the results obtained above, we see clearly that $\rho_{00}$ can be expressed in terms of 
different components of the fragmentation functions. 
We present the results in the following.

From the differential cross section, $\rho_{00}$ can be calculated in the following way,
\begin{align}
\rho_{00} = \frac{d\sigma^{00}}{d\sigma^{++}+d\sigma^{00}+d\sigma^{--}} 
\end{align}
where the superscript of $\sigma$ denotes the helicity of the vector meson. 
These cross sections can easily be calculated by inserting the corresponding values for 
the parameters $S$ into Eqs.~(\ref{vectormeson}) and (\ref{vectormeson2}). 
For example, for $d\sigma^{++}$, we calculate the cross section for vector meson 
in helicity state $\lambda_h=1$ hence, $\rho_{++}=1$ otherwise $\rho_{mm'}=0$. 
This implies that 
$S_{LL}=\frac{1}{2}$, $S_L = 1 $, and all the other components of $S$ are zero. 
Hence, we have,  
\begin{align}
\frac{d\sigma^{++}}{dzdy} = \chi\frac{2\pi\alpha^2}{Q^2} \Bigl\{ T_0 (y) \bigl[ D_1(z) + \frac{1}{2}D_{1LL}(z)\bigr] + T_1(y) \Delta D_{1L} (z)   \Bigr\}.
\end{align}
Integrated over $y$, we have
\begin{align}
\frac{d\sigma^{++}}{dz} = \chi\frac{2\pi\alpha^2}{Q^2} \Bigl\{ t_0 \bigl[ D_1(z)  + \frac{1}{2}  D_{1LL} (z) \bigr] + t_1 \Delta D_{1L} (z) \Bigr\}.
\end{align}
Similarly, for $\lambda_h=0$, $S_{LL}=-1$, $S_L = 0 $, and all the other components of $S$ equal to zero.  Hence, we have, 
\begin{align}
\frac{d\sigma^{00}}{dzdy} =& \frac{2\pi\alpha^2}{Q^2} \chi T_0 (y) \bigl[ D_1(z) -  D_{1LL}(z)  \bigr],\\
\frac{d\sigma^{00}}{dz} = & \frac{2\pi\alpha^2}{Q^2}  \chi t_0 \bigl[ D_1(z)  - D_{1LL} (z) \bigr].
\end{align}
For $\lambda_h=-1$, $S_{LL}=\frac{1}{2}$, $S_L = - 1 $, and other components are zero, so that, 
\begin{align}
& \frac{d\sigma^{--}}{dzdy} = \frac{2\pi\alpha^2}{Q^2} \chi\Bigl\{ T_0 (y) \bigl[D_1(z) + \frac{1}{2}D_{1LL}(z)\bigr]- T_1(y) \Delta D_{1L} (z)   \Bigr\}, \\
& \frac{d\sigma^{--}}{dz} = \frac{2\pi\alpha^2}{Q^2} \chi\Bigl\{ t_0 \bigl[D_1(z)  + \frac{1}{2}  D_{1LL} (z) \bigr]- t_1 \Delta D_{1L} (z) \Bigr\}.
\end{align}
Hence, we obtain $\rho_{00}$ as given by, 
\begin{align}
\rho_{00} (z,y) = \frac{1}{3} - \frac{1}{3} \frac{\sum_{q} T_{0}^{q}(y) D_{1LL}^{q\to h}(z) }{\sum_{q} T_{0}^{q}(y) D_1^{q\to h}(z) }, \label{eq:rhoVres1}
\end{align}
or integrated over $y$ or $\theta$, 
\begin{align}
\rho_{00} (z) = \frac{1}{3} - \frac{1}{3} \frac{\sum_{q} t_{0}^{q} D_{1LL}^{q\to h}(z) }{\sum_{q} t_{0}^{q} D_1^{q\to h}(z) }, \label{eq:rhoVres2}
\end{align}
where we have written out the summation over flavor explicitly.

From the Eqs.~(\ref{eq:rhoVres1}) and (\ref{eq:rhoVres2}), without knowing any detail of the fragmentation functions, 
we are already able to see the following features for the spin alignment parameter $\rho_{00}$ in $e^+e^-$-annihilations. 
First, the spin alignment $\rho_{00}$ for vector mesons produced in $e^+e^-$-annihilations does not depend on 
the polarization of the quark and/or anti-quark produced at the $e^+e^-$-annihilation vertex. 
This can be understood since $\rho_{00}=1-(\rho_{++}+\rho_{--})$ describes only the difference between the 
vector meson in the helicity $\pm 1$ and helicity zero state but has nothing to do with the quark polarization 
in the helicity direction. 
Second, besides the fragmentation function itself, the quark flavor dependence comes in only 
in the relative production weight. 
Since the fragmentation function is determined by strong interaction, the isospin symmetry is valid 
and even SU(3) flavor symmetry is approximately applicable to a good accuracy. 
Furthermore, the spin structures of vector mesons of different flavors are similar to each other. 
Hence, if we consider only the light flavor vector mesons, we expect that $\rho_{00}$ is approximately 
the same for different mesons.  
Such a feature is in contrast to the polarizations for spin-$1/2$ hadrons 
discussed in last sub-section 
where different hyperons are expected to have rather different polarizations. 
This feature for $\rho_{00}$ is consistent with the data available~\cite{Ackerstaff:1997kj, Abreu:1997wd} 
and can be further checked by future experiments. 

For electromagnetic interaction process $e^+e^- \to \gamma^* \to q \bar q \to h +X$,
\begin{align}
\rho_{00}^{\rm em} = \frac{1}{3} - \frac{1}{3} \frac{\sum_{q} e_q^2D_{1LL}^{q\to h}(z)}{\sum_{q} e_q^2D_1^{q\to h}(z)},
\end{align}
which implies that even unpolarized quarks could lead to longitudinally tensor polarized ($S_{LL}$) vector mesons. The qualitative features discussed above apply also here.

\section{The twist-4 contributions}

Unlike the twist-3 contributions, in inclusive hadron production in $e^+e^-$ annihilation 
at high energies, the twist-4 contributions are mostly power suppressed corrections to 
the leading twist contributions whatever measurable quantities that we study.  
Hence, the observable effects led by these twist-4 contributions are usually not very obvious 
and are difficult to separate from the leading twist contributions. 
In this section, we give an example to illustrate how the calculations for such contributions
can be carried out by using the formalism presented in Sec. II. 
We should note that the twist-4 contributions that we present in this section are results from 
the diagram series as illustrated in Fig. 2. 
It is not intend to be a complete study of the twist-4 contributions for the reactions. 
There are also other sources such as four quark correlators that contribute at twist 4. 
A complete study should also take them into account. 
In this section, we only present the results from the diagram series considered in this paper 
to show how to calculate twist-4 contributions in the formulism described in Sec. II. 

From the diagram series that we consider in this paper, the sources of the twist-4 contributions are 
from the quark-quark or quark-gluon-quark correlators such as, 
$\gamma^- \psi \bar\psi $,  $\gamma_\perp \psi D_\perp \bar\psi $, and $ \gamma^+ \psi D_\perp D_\perp \bar\psi $.  
These contributions  are contained in $\tilde W^{(0)}_{\mu\nu} (q,p,S)$, $\tilde W^{(1)}_{\mu\nu} (q,p,S)$ 
and $\tilde W^{(2)}_{\mu\nu} (q,p,S)$ respectively.    
%
%
We can pick them up from Eqs. (\ref{lwsht})-(\ref{t4sht}) and (\ref{t3sp})-(\ref{t4sp}). They are given by,
\begin{align}
&\tilde W^{(0,4)}_{\mu\nu} (q,p,S) = \frac{1}{2} \mathrm{Tr}\left[ \hat h^{(0)}_{\mu\nu} \hat\Xi^{(0)}_-(z_B,p,S;n) \right],\\
&\tilde W^{(1,L,4)}_{\mu\nu} (q,p,S) = - \frac{1}{4p\cdot q} \mathrm{Tr}\left[ \hat h^{(1)\rho}_{\mu\nu}  \omega_\rho^{\ \rho'}  \hat\Xi^{(1)}_{\perp, \rho' } (z_B,p,S;n) \right],\\
&\tilde W^{(2,M,4)}_{\mu\nu} (q,p,S) = \frac{1}{4(p\cdot q)^2}  \mathrm{Tr}\left[ \hat h^{(2)\rho\sigma}_{\mu\nu} \omega_\rho^{\ \rho'} \omega_\sigma^{\ \sigma'} \hat\Xi^{(2A)}_{+\rho'\sigma'}(z_B,p,S;n) \right],\\
& \tilde W^{(2,L,4)}_{\mu\nu} (q,n,S) = \frac{1}{4(p\cdot q)^2}   
\mathrm{Tr} \left[  \hat N^{(2)\rho\sigma}_{\mu\nu} \omega_\rho^{\ \rho'} \omega_\sigma^{\ \sigma'} 
\hat\Xi^{(2C)}_{\rho'\sigma'}(z_B,p,S;n) \right],
\end{align}
and $\tilde W^{(1,R,4)}_{\mu\nu} (q,p,S) = \tilde W^{(1,L,4)*}_{\nu\mu} (q,p,S)$, 
$\tilde W^{(2,R,4)}_{\mu\nu} (q,p,S) = \tilde W^{(2,L,4)*}_{\nu\mu} (q,p,S)$.  
Here we use the number $4$ in the superscript of $\tilde W$ to specify twist-4 contributions. 
The matrices $\hat\Xi^{(0)}_-$, $\hat\Xi^{(1)}_{\perp \rho}$,  and $\hat\Xi^{(2)}_{+ \rho\sigma}$ are 
the ($\gamma_-$, $\gamma_5\gamma_-$)-, ($\gamma_\perp$, $\gamma_5\gamma_\perp$)-, 
and ($\gamma_+$, $\gamma_5\gamma_+$)-components respectively of the corresponding $\hat\Xi$'s. 
They are e.g., defined as,  
$\hat\Xi^{(0)}_-=(\gamma _- \Xi^{(0)}_-+\gamma_5\gamma _- \tilde\Xi^{(0)}_-)$,  
where $\Xi^{(0)}_\alpha$ and $\tilde\Xi^{(0)}_\alpha$ are defined in Eqs (\ref{xi0}) and (\ref{xi05}).
$\hat\Xi^{(0)}_-$ corresponds to the $\gamma^- \psi \bar\psi$ terms, similar for the others.
We pick up these contributions by analysing the Lorentz structures of the corresponding 
$\Xi_\alpha$'s and $\tilde\Xi_\alpha$'s.

%

The Lorentz structure of these components of the corresponding $\Xi$'s that contribute at twist-4 level are given by, 
\begin{align}
&z \Xi^{(0)}_\alpha (z,p,S;n) = \frac{M^2}{p^+}  D_- (z) n_\alpha + ... , \label{twist4start}\\
&z \tilde\Xi^{(0)}_\alpha (z,p,S;n) =\lambda_h \frac{M^2}{p^+}  \Delta D_- (z)  n_\alpha +..., \\
&z \Xi^{(1)\rho\alpha} = i \lambda_h M^2 \varepsilon_\perp^{\rho\alpha} \Delta D_{\perp}^{(1)} (z)
+  M^2 d^{\rho\alpha} D_{\perp}^{(1)} (z) +...,\\
&z \tilde \Xi^{(1)\rho\alpha} = \lambda_h M^2 d^{\rho\alpha} \Delta \tilde D_{\perp}^{(1)} (z) + 
 i M^2 \varepsilon_\perp^{\rho\alpha} \tilde D_{\perp}^{(1)} (z) +  ...,\\
& z \Xi^{(2A)\rho\sigma\alpha} =   i \lambda_h  M^2 \varepsilon_\perp^{\rho\sigma} p^\alpha \Delta D^{(2)}(z) + 
M^2 d^{\rho\sigma} p^\alpha D^{(2)}(z) +...,\\
& z\tilde{\Xi}^{(2A)\rho\sigma\alpha} =  \lambda_h M^2 d^{\rho\sigma} p^\alpha \Delta \tilde D^{(2)}(z) 
+ i  M^2 \varepsilon_\perp^{\rho\sigma} p^\alpha \tilde D^{(2)}(z) +....\\
& z \Xi^{(2C)\rho\sigma\alpha} =   M^2 d^{\rho\sigma}  p^\alpha D^{(2L)} (z)+ 
i\lambda_h M^2\varepsilon_\perp^{\rho\sigma}  p^\alpha \Delta D^{(2L)} (z) ,\\
& z \tilde \Xi^{(2C)\rho\sigma\alpha} = 
M^2 d^{\rho\sigma} \lambda_h  p^\alpha \Delta \tilde D^{(2L)} (z) + 
i  M^2 \varepsilon_\perp^{\rho\sigma}  p^\alpha \tilde D^{(2L)} (z). \label{twist4end}
\end{align}
Here, the subscript of the $D$'s or $\tilde D$'s to specify that it comes from $\bar n$, $\perp$ or $n$-component, 
the superscript specifies from which $\Xi$ it originates; those with $\Delta$ are longitudinal spin dependent, and 
those without $\Delta$ are spin independent. We see that the $D_-(z)$-term just corresponds to 
the $n$-component of the hadron momentum $p$ as we mentioned in Sec. II. 

Again, equation of motion $\gamma\cdot D\psi(z)=0$ relates,  
\begin{align}
\frac{1}{z^2}  D_- (z)&=\frac{1}{z}\left[ D_{\perp}^{(1)} (z)- \tilde D_{\perp}^{(1)}  (z)\right] 
= -\left[ D^{(2)} (z) + \tilde D^{(2)} (z) \right] ,\label{eom4s}\\
\frac{1}{z^2} \Delta D_- (z) &=\frac{1}{z}\left[ \Delta \tilde D_\perp^{(1)}  (z)- \Delta D_\perp^{(1)} (z) \right] 
= -\left[ \Delta \tilde D^{(2)} (z) + \Delta D^{(2)} (z) \right].  \label{eom4e}
\end{align}

By inserting Eqs.~(\ref{twist4start})-(\ref{twist4end}) into Eqs.~(\ref{lwsht})-(\ref{t4sht}) 
and carrying out the traces and simplifying the results using Eqs.~(\ref{eom4s})-(\ref{eom4e}), 
we obtain the final twist-4 contributions to the hadronic tensor,
\begin{align}
W^{(4)}_{\mu\nu}(q,p,S)
& = \frac{16M^2}{z^3Q^4} \Bigl\{ (q-2p/z)_\mu(q-2p/z)_\nu [ c_1^q D_-(z)- \lambda_h c_3^q \Delta D_-(z) ]\nonumber\\
& - \frac{z^2}{4} Q^2  [(c_1^q d_{\mu\nu} + i c_3^q\varepsilon_{\perp\mu\nu} ) D_4^{(2L)}(z) + 
\lambda_h (c_3^q d_{\mu\nu} + ic_1^q \varepsilon_{\perp\mu\nu} ) \Delta D_4^{(2L)}(z)] \Bigr\},  
\end{align}
where the new symbols $D_4^{(2L)}$ and $\Delta D_4^{(2L)}$ are defined as,
\begin{align}
&zD_4^{(2L)} (z) \equiv \mathrm{Re}[\tilde D^{2L} (z) - D^{(2L)} (z)], \\
& z\Delta D_4^{2L} (z) \equiv \mathrm{Re} [\Delta D^{(2L)} (z) - \Delta \tilde D^{(2L)} (z)].
\end{align}

After making contraction with the leptonic tensor $L_{\mu\nu}$, we obtain 
the twist-4 contributions to the cross section as,
\begin{align}
E_p\frac{d\sigma}{d^3p} =  \frac{8\alpha^2M^2}{Q^6z^3}\chi
\Bigl\{ & [T_4 (y) D_- (z) +T_0(y) z^2 D_4^{(2L)} (z)   ] \nonumber \\
 + &\lambda_h [-  T_5 (y) \Delta D_- (z) + T_1 (y) z^2 \Delta D_4^{(2L)} (z) ] \Bigr\}, 
\end{align}
where the two new coefficient functions of $y$ are given by, 
\begin{align}
& T_4^q(y)= 4y(1-y) c_1^e c_1^q=\frac{|\vec l_\perp|^2}{Q^2}\frac{d^2 T_0(y)}{dy^2},\\
& T_5^q(y)= 4y(1-y) c_1^e c_3^q=-\frac{|\vec l_\perp|^2}{Q^2}\frac{d^2 T_1(y)}{dy^2}.
\end{align}
We note that $T_4^{\bar q}(y) = T_4^q(y)$ and $T_5^{\bar q}(y) = T_5^q(y)$. 
We see that there are terms that contribute to the unpolarized cross section and those to the longitudinal polarization.
Up to twist-4 level, we should add these contributions to the leading twist contributions to obtain the final results. 
However, for the observables such as the production rates, the spectra and the longitudinal polarizations, 
these contributions are just higher twist addenda suppressed by the factor $M^2/Q^2$ and in general are difficult to be separated from the leading contributions.

\section{Summary and outlook}

In summary, we apply the collinear expansion to inclusive hadron production in $e^+e^-$-annihilations at high energies. 
We derive the formalism that can be used to study the leading as well as higher twist contributions in a systematic and 
consistent way. We calculate the contributions to the production of hadrons with different spins up to twist-3 level. 
We also present the results for spin-1/2 hadrons at the twist-4 level. 
The results clearly show a number of interesting features. 
In the unpolarized case or for spin-zero hadrons, the cross section has the expression as usually used. 
For hadron with spins, there are leading twist longitudinal polarization for spin-1/2 hadrons in 
$e^+e^-\to Z\to q\bar q\to h+X$ because the initial quark and anti-quark produced here are longitudinally polarized 
and such polarizations can be transferred to the hadrons produced.  There is also spin alignment $\rho_{00}\not =1/3$ 
for spin-1 i.e. vector mesons, and the spin alignment is independent of the polarization of the initial quark or anti-quark 
thus exist also in $e^+e^-\to \gamma^* \to q\bar q\to h+X$.  

At the twist-3 level, there is a transverse polarization of spin-$1/2$ hadrons 
in the leptonic plane as well as transverse to the leptonic plane. 
The component of such transverse polarization in the leptonic plane vanishes in  
$e^+e^-\to \gamma^* \to q\bar q\to h+X$ but the component transverse to 
the leptonic plane still remains. 

In inclusive hadron production in $e^+e^-$ annihilation at high energies, 
twist-4 contributions are usually power suppressed 
addenda to leading twist contributions and do not lead to new observable effects.

The formalism should also be extended to semi-inclusive hadron production process 
where transverse momentum dependent fragmentation functions can also be studied. 
Such a study is underway.

\section*{Acknowledgements}
This work is supported in part by the National Natural Science Foundation of China projects (Nos. 11035003 and 11375104) 
and the Major State Basic Research Development Program in China (No. 2014CB845400). 
YKS was supported in part by CCNU-QLPL Innovation Fund (QLPL 2011P01).


\end{document}